\documentclass[11pt,a4paper]{article}
\usepackage{amsmath}
\usepackage{cite}
\usepackage{graphicx}
\usepackage{amsfonts}
\usepackage{amssymb}

\def\clebsch#1#2#3#4#5#6{
\left\lbrack
\begin{array}
[c]{ccc}%
#1 & #2 & #3\\
#4 & #5 & #6%
\end{array}\right\rbrack}

\def\my6j#1#2#3#4#5#6{
\left\{
\begin{array}
[c]{cc}%
#1 & #3\\
#2 & #4%
\end{array}
\mid
\begin{array}{c}
#5\\
#6%
\end{array}\right\}_{b}}

\def\fus#1#2#3#4#5#6{
F_{#5#6}\left[
\begin{array}
[c]{cc}%
#3 & #2\\
#4 & #1%
\end{array}
\right]}

\def\gfus#1#2#3#4#5#6{
G_{#5#6}\left[
\begin{array}
[c]{cc}%
#3 & #2\\
#4 & #1%
\end{array}
\right]}

\def\phis#1#2#3#4#5{
{\Phi}^{s}_{#5}\left[
\begin{array}
[c]{cc}%
#3 & #2\\
#4 & #1%
\end{array}
\right](x)}

\def\thetas#1#2#3#4#5{
{\Theta}^{s}_{#5}\left[
\begin{array}
[c]{cc}%
#3 & #2\\
#4 & #1%
\end{array}
\right](x)}

\def\psis#1#2#3#4#5{
{\Psi}^{s}_{#5}\left[
\begin{array}
[c]{cc}%
#3 & #2\\
#4 & #1%
\end{array}
\right](x)}

\def\thetat#1#2#3#4#5{
{\Theta}^{t}_{#5}\left[
\begin{array}
[c]{cc}%
#3 & #2\\
#4 & #1%
\end{array}
\right](x)}

\def\phit#1#2#3#4#5{
{\Phi}^{t}_{#5}\left[
\begin{array}
[c]{cc}%
#3 & #2\\
#4 & #1%
\end{array}
\right](x)}

\def\psit#1#2#3#4#5{
{\Psi}^{t}_{#5}\left[
\begin{array}
[c]{cc}%
#3 & #2\\
#4 & #1%
\end{array}
\right](x)}

\def\norms#1#2#3#4#5{
N^{s}_{#5}\left[
\begin{array}
[c]{cc}%
#3 & #2\\
#4 & #1%
\end{array}\right]}

\def\normt#1#2#3#4#5{
N^{t}_{#5}\left[
\begin{array}
[c]{cc}%
#3 & #2\\
#4 & #1%
\end{array}\right]}

\def\Sp{
\mathbb{S} }

\def\al{
\alpha }

\def\Ga{
\Gamma_{b} }

\def\up{\Upsilon_{b}}

\def\del{\delta_{x}}
\def\de{\Delta}

\def\p#1{\mathcal{P}_{\al _{#1}}}
\def\uq{\mathcal{U}_q(\mathfrak{sl}(2,\mathbb{R}))}

\begin{document}
\begin{center}
{\Large \textbf{Recent progress in Liouville field theory\footnote{This article gathers results
 obtained in collaboration with J.~Teschner \cite{PT1,PT2,PT3} between 1997 and 2001.} }}\\
\vspace{1.5cm} {\Large B\'en\'edicte Ponsot } \\
\vspace{0.7cm} {\it  Service de Physique Th\'eorique, Commissariat
\`a l'\'energie
atomique,\\
 CEA L'Orme des Merisiers,
 F-91191 Gif sur Yvette, France.}
\end{center}
\vspace{0.5cm}
\begin{abstract}
An explicit construction for the monodromy of the Liouville
conformal blocks in terms of Racah-Wigner coefficients of the quantum group
$\uq$ is
proposed. As a consequence, crossing-symmetry for four point functions is analytically proven, and the expression for the
correlator of three boundary operators is obtained.
\end{abstract}

\section{Introduction}
Liouville field theory has permanently attracted attention in the
last two decades in the context of strings in non critical
space-time dimensions; early works on the subject can be found in
 \cite{Polyakov,curtright,DJ,DJ2,DFJ,GN}. It is also an
alternative approach to matrix models in the study of 2D gravity
\cite{Kazakov}. In the context of the $AdS$/CFT correspondence
(see \cite{Malda} and references therein), it is also very closely
related to the $SL(2,\mathbb{C})/SU(2)$ and $SL(2,\mathbb{R})$
WZNW models that describe strings propagating on Euclidean and
 Lorentzian $AdS_3$.\\ Liouville
field theory is the simplest case of non compact conformal field
theories, i.e. with a continuous spectrum of primary fields, and
serves as a prototype to develop techniques that can also be
useful in the study of more complicated CFT that share similar
features such as non compactness.\\
Given knowledge of conformal symmetry, spectrum and three point
function, one has an unambiguous construction for any genus zero
correlation function by summing over intermediate states. But the
decomposition of an n-point function as a sum over three point
functions can be performed in different ways. Equivalence of such
decompositions (crossing symmetry, also called bootstrap condition) can be seen as the most
difficult sufficient condition to verify in showing consistency of
the CFT as characterized by the spectrum and the three point
function. In this lecture, we will first review the articles \cite{PT1,PT2}. In \cite{PT1}, the
crossing-symmetry condition for four-point functions is proven analytically in the weak coupling regime
 where the Liouville central charge $c_L \geq 25$, thanks to
the explicit construction of the monodromy (or fusion matrix) of
the conformal blocks\footnote{Another method obtained after
 the one presented here can be found in
\cite{T}.}. The expressions can be continued into the strong coupling regime $1<c_L<25$ by analytical
 continuation, providing thus a solution of the bootstrap in this region. The fusion matrix is constructed
  in terms of Racah-Wigner coefficients (or $6j$ symbol)
for an appropriate continuous series of representations of the quantum group $\uq$,
where the deformation parameter $q=e^{i\pi b^2}$ ($b$ real) is related to
the Liouville central charge by
$c_L=1+6(b+b^{-1})^2,\; c_L \geq 25$. We refer the reader to \cite{PT2}
for the mathematical details of this construction.\\
Although the relation between CFTs and quantum groups
 is still rather mysterious,
there is a long story of connections between Liouville field
theory and quantum groups \cite{Takh,Takh2,bab,G}, and more
recently in \cite{GS,CGR,GR}. Our approach is somewhat an
extension of the one developed in the latter papers. We would like
to mention that the representations of $\uq$ whose study is
relevant for our purposes (we also refer the
 reader to \cite{Faddeev} for discussions on the closely related concept of modular double,
  first introduced by L.~Faddeev), are continuous,
autodual in the sense that they remain unchanged when $b$ is replaced
 by $b^{-1}$, and do not have any
classical counterpart.\\
Once the monodromy of the conformal blocks
is constructed, it is not so difficult to deduce the explicit
expression for the boundary three point function: an ansatz for it
in terms of the fusion matrix leads to a solution of the
consistency condition that expresses the associativity of the
product of boundary operators \cite{R}. The last task is to fix the
remaining freedom by imposing certain normalization conditions.
This result is published in \cite{PT3} and is derived in the
last section of the lecture.

\section{Preliminaries}
We consider Liouville field theory defined in the bulk by the
following action:
\begin{eqnarray}
\mathcal{A}_{L}=\frac{1}{4\pi}\int_{\Gamma}\left[g^{ab}\partial_a\phi\partial_b\phi+QR\phi
+ 4\pi\mu e^{2b\phi}\right]\sqrt{g}d^{2}x \; ,\nonumber
\end{eqnarray}
where $R$ is the scalar curvature associated with the background metric
$g$ and $Q=b+1/b$ is called the background charge that determines the
central charge of the theory
$$
c_L=1+6Q^2\; .
$$
We will consider in the following the so-called weak coupling
regime, where the central charge $c_L \geq 25$. Let us note that
the latter remains unchanged under the transformation $b \to 1/b$;
so we take here the opportunity to mention that at the quantum
level ({\it only}), Liouville theory enjoys the property of {\it
autoduality}, i.e. the observables of
the theory are invariant when $b$ and $1/b$ are exchanged.\\
In what follows, we will consider Liouville theory on the flat
plane with trivial background metric $g_{ab}=\delta_{ab}$; in this
case the action reads
\begin{eqnarray}
\mathcal{A}_{L}=\int\left(\frac{1}{4\pi}(\partial_a\phi)^{2}+\mu
e^{2b\phi}\right)d^{2}x \; ,\label{action Liou}\nonumber
\end{eqnarray}
with the curvature sent at spatial infinity, which means for the Liouville
field the following boundary condition
\begin{equation}
\phi(z,\bar{z})=-Q\log(z\bar{z})+O(1)\ \ \ \ \ \mathrm{at}\ \
|z|\rightarrow \infty.\nonumber
\end{equation}
We note the conformal primary fields $V_{\al}(z,\bar{z})$
(classically one has $V_{\al}(z,\bar{z})=e^{\al/b \varphi(z,\bar{z})}$,
where $\varphi=2b\phi$ is the classical Liouville field). These fields
 are
primaries with respect to the energy momentum tensor
\begin{eqnarray}
T(z)&=&-(\partial\phi)^{2}+Q\partial^{2}\phi \; ,\nonumber \\
\bar{T}(\bar{z})&=&-(\bar{\partial}\phi)^{2}+Q\bar{\partial}^{2}\phi \; .
\nonumber
\end{eqnarray}
and have conformal weight
$\Delta_{\alpha}=\bar{\Delta}_{\alpha}=\alpha(Q-\alpha)$. Because
of the invariance $\al \to Q-\al$, one identifies the operator
$V_{\al}$ with its reflected image $V_{Q-\al}$. In this lecture we
will consider the following two important sets of values for
$\al$.
\begin{itemize}
\item Space of states: $\al=\frac{Q}{2}+iP,\; P>0$\\
Because of conformal invariance, the space of states $\mathcal{H}$
decomposes as a direct sum over tensor products
$\mathcal{V}_{\al}\otimes \overline{\mathcal{V}}_{\al}$ of highest
weight representations of the left/right Virasoro algebra.
Arguments based on canonical quantization \cite{curtright} suggest
the following spectrum:
\begin{equation}
\mathcal{H}=\int_{\Sp}^{\oplus}\mathcal{V}_{\al}\otimes
\overline{\mathcal{V}}_{\al},\, \quad
\Sp=\frac{Q}{2}+i\mathbb{R}^+\nonumber
\end{equation}
where each $\mathcal{V}_{\al}$ contains a primary state $v_{\al}$
satisfying
\begin{eqnarray}
L_nv_{\al}&=&\bar{L}_nv_{\al}=0,\quad n>0\nonumber \\
L_{0}v_{\al}&=&\bar{L}_{0}v_{\al}=\left(Q^{2}/4+P^{2}\right)v_{\al}\; .
\nonumber
\end{eqnarray}
\item Degenerate Virasoro representations:\\
The fields $V_{-nb/2},\;n \in \mathbb{N},$ are degenerate with
respect to the conformal symmetry algebra and satisfy linear
differential equations \cite{BPZ}. For example, the first non trivial case
consists of $\al=-b/2$, and the corresponding operator satisfies
\begin{equation}
\left(\frac{1}{b^{2}}\partial^{2}+T(z)\right)V_{-b/2}=0, \nonumber
\end{equation}
as well as the complex conjugate equation.\\
For $n=2$, the operator $V_{-b}$ satisfies two complex conjugate
differential equations of the third order
\begin{equation}
\left(\frac{1}{2b^{2}}\partial^{3}+2T(z)\partial+(1+2b^{2})\partial
T(z)\right)V_{-b}=0. \nonumber
\end{equation}
It follows from these equations that when one performs the
operator product expansion of one of these degenerate operators
with a generic operator, then the OPE is truncated \cite{BPZ}.
For example:
\begin{eqnarray}
V_{-b/2}V_{\al}=c_{+}V_{\al-b/2}+c_{-}V_{\al+b/2} \, .\label{OPE}
\nonumber
\end{eqnarray}
So $\al=-b/2$ is nothing but the usual spin $1/2$ (multiplied by
$-b$); according to the fusion rules, its tensor product with a
generic representation of spin $\al_1$ gives a sum of
representations with spin $\al_{21}=\al_1\pm b/2$. The structure
constants $c_{\pm}$ are special cases of the bulk three point
function, and can be computed perturbatively\footnote{One of the first step in this
direction was developed in \cite{GL}.} as Coulomb gas (or
screening) integrals \cite{FF,DF}. One can take $c_+=1$, as in
this case there is no need of insertion of interaction, whereas
$c_-$ requires one insertion of the Liouville interaction
$-\mu\int e^{2b\phi}d^2z $, and
\begin{eqnarray}
c_-&=&-\mu\int d^2z \left\langle V_{\alpha}(0)
V_{-b/2}(1)e^{2b\phi(z,\bar{z})}V_{Q-\alpha-b/2}(\infty)\right\rangle\nonumber
\\
&=& -\mu\int d^2z |z|^{2b\al}|1-z|^{-b^2}\nonumber \\
&=&
-\mu\frac{\pi\gamma(2b\alpha-1-b^2)}{\gamma(-b^2)\gamma(2b\al)} \, .\nonumber
\end{eqnarray}
where $\gamma(x)=\Gamma(x)/\Gamma(1-x)$; in the first line, we
used the property of invariance under global transformations to
set $z_1=0,\, z_3=1,\, z_4=\infty$, and in the second line
$<\phi(x)\phi(y)>=-\log |x-y|$.\\
There exists also a dual series of degenerate operators
$V_{-m/2b}$ which have truncated operator expansion. If $\alpha_2$
is a degenerate Virasoro representation ($\al_2=-nb/2-m/2b$), then
the fusion rules are such that $\al_{21}-\al_1-\al_2=sb+tb^{-1},\;
s,t\in \mathbb{N}$ \cite{BPZ}. In this case, the structure constants can be
computed pertubatively with $s$ insertions of the Liouville
interaction $-\mu\int e^{2b\phi(z,\bar{z})}d^2z $ and $t$
insertions of the {\it dual} interaction $-\tilde{\mu}\int
e^{2b^{-1}\phi(z,\bar{z})}d^2z$ where the dual cosmological
constant  $\tilde{\mu}$ is related to the cosmological constant by
the formula
\begin{eqnarray}
\pi\tilde{\mu}\gamma(1/b^2)=(\pi\mu\gamma(b^2))^{1/b^2}.\label{mu}
\end{eqnarray}
This means that from the point of view of the path integral,
everything occurs as if we had added a dual interaction term in
the action. This term is compatible with conformal invariance, as
$e^{2b^{-1}\phi}$ has conformal weight $(1,1)$. Finally, if
$\al_{21}-\al_1-\al_2=sb+tb^{-1}$, then the three point function
can be evaluated by computing the integral
\begin{eqnarray}
\lefteqn{\frac{(-\mu)^{s}(-\tilde{\mu})^{t}}{s!t!}\int d^2u_1\dots
d^2u_s
\int d^2v_1\dots d^2v_t} \nonumber \\
&& \left\langle
V_{\alpha_1}(0)V_{\al_2}(1)\prod_{i=1}^{s}e^{2b\phi(u_i,\bar{u}_i)}
\prod_{j=1}^{t}e^{2b^{-1}\phi(v_j,\bar{v}_j)}
V_{Q-\alpha_{21}}(\infty)\right\rangle \; .\nonumber
\end{eqnarray}
whose explicit expression can be found in \cite{DF}.
\end{itemize}
\subsection{Three point function \cite{Dorn,ZZ}}
The spatial dependence of the three point correlation function is
completely determined by conformal symmetry,
$$
\mathcal{G}_{\al_1,\al_2,\al_3}(z_1,z_2,z_3)=
|z_{12}|^{2\gamma_3}|z_{32}|^{2\gamma_1}|z_{31}|^{2\gamma_2}C(\al_3,\al_2,\al_1)
$$
where
$\gamma_1=\Delta_{\al_1}-\Delta_{\al_2}-\Delta_{\al_3},\gamma_2=\Delta_{\al_2}-\Delta_{\al_3}-\Delta_{\al_1},
\gamma_3=\Delta_{\al_3}-\Delta_{\al_1}-\Delta_{\al_2}$. It remains
only to determine the pure number $C(\al_3,\al_2,\al_1)$. For
generic values of $\al_i,\; i=1,2,3$, the following expression for
this structure constant was proposed by \cite{Dorn,ZZ}
\begin{eqnarray}
\lefteqn{C(\al_3,\al_2,\al_1)= \left\lbrack \pi \mu \gamma(b^{2})b^{2-2b^{2}}
\right\rbrack^{\frac{Q-\al_1-\al_2-\al_3}{b}}}. \nonumber \\
&&
\frac{\Upsilon_0\up(2\al_1)\up(2\al_2)\up(2\al_3)}{\up(\al_1+\al_2+\al_3-Q)
\up(\al_1+\al_2-\al_3)\up(\al_1+\al_3-\al_2)\up(\al_2+\al_3-\al_1)}
\nonumber \label{3pointsZZ}
\end{eqnarray}
where the special function\footnote{Definitions and properties of
the special functions $\Gamma_b,\up,S_b$ are collected in the
Appendix B.} $\up$ is such defined: $\up^{-1}(x)\equiv
\Ga(x)\Ga(Q-x)$, with $\Gamma_b(x)\equiv
\frac{\Gamma_2(x|b,b^{-1})}{\Gamma_2(Q/2|b,b^{-1})}$, and
$\Gamma_2(x|\omega_1,\omega_2)$ is the Double Gamma function
introduced by Barnes \cite{Barnes}, which definition is
\begin{eqnarray}
\nonumber \\
&&
\text{log}\Gamma_{2}(x|\omega_1,\omega_2)=\left(\frac{\partial}{\partial
t}\sum_{n_1,n_2=0}^{\infty}(x+n_1\omega_1+n_2\omega_2)^{-t}\right)_{t=0}
\; .\nonumber
\end{eqnarray}
The function $\Ga(x)$ defined above is such that
$\Gamma_b(x)\equiv\Gamma_{b^{-1}}(x)$, and satisfies the following
functional relation
\begin{eqnarray}
\Ga(x+b)=
\frac{\sqrt{2\pi}b^{bx-\frac{1}{2}}}{\Gamma(bx)}\Ga(x)\, ,\nonumber
\end{eqnarray}
as well as the dual functional relation with $b$ replaced by
$1/b$. $\Upsilon_{0}=\text{res}_{x=0}\frac{d\up(x)}{dx}$.\\
 The
three point function enjoys remarkable properties:
\begin{itemize}
  \item
It has poles located at $\al_{21}-\al_1-\al_2=sb+tb^{-1},\; s,t
\in\mathbb{N}$ (other poles are obtained by reflection $\al_i \to
Q-\al_i$), whose residues exactly coincide with the pertubative
computation in terms of screening integrals explained above.

  \item
It satisfies the reflection property, for each $\al_i$
\begin{eqnarray}
C(\al_3,\al_2,\al_1)=C(\al_3,\al_2,Q-\al_1)S(\al_1)\, , \label{ral}
\end{eqnarray}
where $S(\al)$ is the reflection amplitude that enters the two
point function
\begin{eqnarray}
\lefteqn{\left\langle V_{\al_{2}}(z_2,\bar{z}_2)V_{\al_1}(z_1,\bar{z}_1)\right\rangle =} \nonumber \\
&&
\left[\delta(\al_{2}+\al_{1}-Q)+S(\al_1)\delta(\al_{2}-\al_{1})\right]|z_1-z_2|^{4\Delta_{\al_1}}\; .
\nonumber  \label{2pointsbulk}
\end{eqnarray}
Explicitly:
$$
S(\al)=\frac{(\pi \mu
\gamma(b^{2}))^{(Q-2\al)/b}}{b^{2}}\frac{\gamma(2\al
b-b^{2})}{\gamma(2-2\al/b+1/b^{2})} \, .\label{amplitude bulk}
$$
The reflection amplitude satisfies the unitarity property $$S(\al)S(Q-\al)=1.$$
 \item
It is invariant when $b$ is replaced by $1/b$, and $\mu$ replaced
by $\tilde{\mu}$ defined in (\ref{mu}).
\end{itemize}

\subsection{Crossing symmetry}
With the three point function and the space of states at hand,
we are in a position to write down a four point function by
summing over intermediate states. Let us denote the four point
function
\begin{eqnarray}
\left\langle
V_{\al_4}(\infty)V_{\al_3}(1)V_{\al_{2}}(z,\bar{z})V_{\al_1}(0)\right\rangle
\equiv \mathcal{V}_{\al_4,\al_3,\al_2,\al_1}(z,\bar{z})\; . \nonumber
\end{eqnarray}
It can be written first in the $s$-channel
\begin{eqnarray}
\mathcal{V}_{\al_4,\al_3,\al_2,\al_1}(z,\bar{z})=
\int_{\mathcal{D}}d\al_{21}C(\al_4,\al_3,\al_{21})C(Q-\al_{21},\al_{2},\al_{1})
|\mathcal{F}^{s}(\Delta_{\al_i},\Delta_{\al_{21}}|z)|^{2}
\nonumber
\end{eqnarray}
\begin{itemize}
\item
$\mathcal{F}^{s}(\Delta_{\al_i},\Delta_{\al_{21}}|z)$ is the
$s$-channel Liouville {\it conformal block} \cite{BPZ},
represented by power series of the form
\begin{eqnarray}
\mathcal{F}^{s}(\Delta_{\al_i},\Delta_{\al_{21}}|z)=
 z_{43}^{\Delta_2+\Delta_1-\Delta_3-\Delta_4}
z_{42}^{-2\Delta_2}z_{41}^{\Delta_3+\Delta_2-\Delta_1-\Delta_4}z_{31}^{\Delta_4-\Delta_1-\Delta_2-\Delta_3}\nonumber
\\
 \times
z^{\Delta_{21}-\Delta_1-\Delta_2}\sum_{n=0}^{\infty}z^n\mathcal{F}^{s}_n(\Delta_{\al_i},\Delta_{\al_{21}})\nonumber
\end{eqnarray}
where $z_{ij}=z_i-z_j,\; i,j=1,\dots ,4$ and the cross ratio
$z=\frac{(z_1-z_2)(z_3-z_4)}{(z_1-z_3)(z_2-z_4)}$. It is a {\it
chiral} object, completely determined by Virasoro symmetry
(although there is no closed form known for it in general). It
depends on conformal weights only. We can use invariance under
global transformations to set $z_1=0, z_3=1, z_4=\infty$, so
$z\equiv z_2$. The conformal block is normalized such that
\begin{eqnarray}
\mathcal{F}^{s}(\Delta_{\al_i},\Delta_{\al_{21}},z)\sim_{z\rightarrow
0}z^{\de_{\al_{21}}-\de_{\al_1}-\de_{\al_2}}(1+\mathcal{O}(z)).\nonumber
\end{eqnarray}
\item
We would like to stress that in general, the set $\mathcal{D}$
does not coincide with the spectrum $\Sp$. This is the case
however if the external $\al_i$ are in the range
$$
\begin{array}{ll}
 2|\text{Re}(\al_1+\al_2-Q)|<Q,\quad     &2|\text{Re}(\al_1-\al_2)|<Q, \\
 2|\text{Re}(\al_3+\al_4-Q)|<Q,\quad     &2|\text{Re}(\al_3-\al_4)|<Q. \\
\end{array}
$$
as the three point functions are analytic in this range. Outside
of this range, meromorphic continuation should be understood. Some poles
of the three point functions will cross the real $P$ axis and
their residues will give additional contributions besides the
integral over the spectrum \cite{Pol,ZZ}. For our discussion, it
will be enough to consider the external spins in the range above,
so we will integrate over the spectrum only.
  \item The four point function
can be written equivalently in the $t$-channel
\begin{eqnarray}
\lefteqn{\mathcal{V}_{\al_4,\al_3,\al_2,\al_1}(z,\bar{z})=}\nonumber
\\
&&
\int_{\Sp}d\al_{32}C(\al_4,\al_{32},\al_{1})C(Q-\al_{32},\al_{3},\al_{2})
|\mathcal{F}^{t}(\Delta_{\al_i},\Delta_{\al_{32}}|1-z)|^{2}
\nonumber
\end{eqnarray}
By comparison with the $s$-channel, $(\al_1,z_1)$ and
$(\al_3,z_3)$ have been exchanged, so the cross ratio $z$ is
replaced by $1-z$ in the $t$-channel.
   \item
There exist\footnote{This is supported by an explicit computation in
a particular case treated in \cite{N}, and was subsequently proven
in \cite{T}.} invertible fusion transformations between $s$- and
$t$-channel conformal blocks, defining the Liouville fusion matrix
(or monodromy of conformal blocks):
\begin{eqnarray}
\mathcal{F}^{s}(\Delta_{\al_i},\Delta_{\al_{21}}|z)=\int_{\Sp}
d\al_{32}\fus{\al_1}{\al_2}{\al_3}{\al_4}{\al_{21}}{\al_{32}}\mathcal{F}^{t}(\Delta_{\al_i},\Delta_{\al_{32}}|1-z) \, .
\label{transfost}\nonumber
\end{eqnarray}
Then the crossing symmetry condition becomes
\begin{eqnarray}
\int_{\Sp}d\al_{21}C(\al_4,\al_3,\al_{21})C(Q-\al_{21},\al_{2},\al_{1})
\fus{\al_1}{\al_2}{\al_3}{\al_4}{\al_{21}}{\al_{32}}
\left(\fus{\al_1}{\al_2}{\al_3}{\al_4}{\al_{21}}{\beta_{32}}\right)^* \nonumber\\
=\delta(\al_{32}-\beta_{32})C(\al_4,\al_{32},\al_{1})C(Q-\al_{32},\al_{3},\al_{2})\nonumber
\end{eqnarray}
\end{itemize}
\subsubsection{Properties of the fusion matrix}
\begin{itemize}
  \item
  It is quite obvious that the fusion matrix should be invariant
   when exchanging columns and rows:
\begin{eqnarray}
\fus{\al_1}{\al_2}{\al_3}{\al_4}{\al_{21}}{\al_{32}}=
\fus{\al_2}{\al_1}{\al_4}{\al_3}{\al_{21}}{\al_{32}}=
\fus{\al_4}{\al_3}{\al_2}{\al_1}{\al_{21}}{\al_{32}}\nonumber
\end{eqnarray}
  \item
  As the conformal blocks depend on conformal weights only, so does the fusion matrix, i.e. it remains
   unchanged
  when any of the six $\al_i,\; i=1,\dots,4,(21),(32),$ is changed into
  $Q-\al_i$.
  \item Moore-Seiberg equations (or polynomial equations) \cite{MS}.\\
Although the Moore-Seiberg equations are proven to hold in
rational conformal field theory, we believe they continue to hold
even in the non compact case for coherency of the operator
algebra. Actually, we will prove in the next section that they
{\it do} hold in the case of Liouville field theory. Let us
remind that they consist of
\begin{itemize}
  \item one pentagonal equation
  \begin{align}
\int_{\frac{Q}{2}+i\mathbb{R^{+}}}d\delta_{1}\fus{\al_1}{\al_2}{\al_3}{\beta_{2}}{\beta_{1}}{\delta_{1}}
\fus{\al_1}{\delta_1}{\al_4}{\al_5}{\beta_{2}}{\gamma_{2}}
\fus{\al_2}{\al_3}{\al_4}{\gamma_{2}}{\delta_{1}}{\gamma_{1}}\nonumber \\
=\fus{\beta_1}{\al_3}{\al_4}{\al_5}{\beta_{2}}{\gamma_{1}}
\fus{\al_1}{\al_2}{\gamma_1}{\al_{5}}{\beta_{1}}{\gamma_{2}}\label{pentagone}
\end{align}
  \item two hexagonal equations
  \begin{align}
\lefteqn{\fus{\al_1}{\al_2}{\al_4}{\al_3}{\al_{21},}{\beta}e^{i\pi\epsilon(\de_{\al_1}+\de_{\al_2}+\de_{\al_3}+\de_{\al_4}-\de_{\al_{21}}-\de_{\beta)}}=} \nonumber \\
&&
\int_{\frac{Q}{2}+i\mathbb{R}^{+}}d\al_{32}\fus{\al_1}{\al_2}{\al_3}{\al_4}{\al_{21},}{\al_{32}}\fus{\al_1}{\al_4}{\al_2}{\al_3}{\al_{32},}{\beta}e^{i\pi
\epsilon\Delta_{\al_{32}}}\nonumber
\end{align}
where $\epsilon =\pm$.\\
\end{itemize}

These equations generate in genus zero all other polynomial equations
 \cite{MS}.
\item
If one of the $\al_i,\; i=1,\dots,4$, (say $\al_2$), corresponds
to a degenerate Virasoro representation, then the fusion
transformations simplify: the conformal blocks then exist only for
a {\it finite} number of values of $\al_{21}$ and $\al_{32}$, so
that the fusion coefficients form a finite dimensional matrix
\cite{BPZ}. It is enough to consider the case where $\al_2=-b/2$
or $-1/2b$, higher cases are generated by the use of the
pentagonal equation\footnote{For example, the fusion coefficients
$\fus{-b}{\al_3}{\al_4}{\al_5}{\beta_{2}}{\gamma_{1}}$ are
computed by setting in the pentagonal equation
$\al_1=\al_2=-b/2$.}. For $\al_2=-b/2$, the conformal blocks are
known explicitly in terms of hypergeometric functions $_2F_1$
\cite{BPZ}. They read in the $s$-channel:
\begin{eqnarray}
\mathcal{F}^s_{+}&=&z^{b\al_1}(1-z)^{b\al_3}\times\nonumber\\
&&
_2F_1(b(\al_1+\al_3+\al_4-3b/2)-1,b(\al_1+\al_3-\al_4-b/2);b(2\al_1-b);z)\, ,\nonumber\\
\mathcal{F}^s_{-}&=&z^{1-b(\al_1-b)}(1-z)^{b\al_3}\times\nonumber\\
&&_2F_1(b(-\al_1+\al_3+\al_4-b/2),b(-\al_1+\al_3-\al_4+b/2)+1;2-b(2\al_1-b);z)
\, .\nonumber
\end{eqnarray}
and in the $t$-channel:
\begin{eqnarray}
\mathcal{F}^t_{+}&=&z^{b\al_1}(1-z)^{b\al_3}\times\nonumber \\
&&_2F_1(b(\al_1+\al_3+\al_4-3b/2)-1,b(\al_1+\al_3-\al_4-b/2);b(2\al_3-b);1-z)\nonumber \, ,\\
\mathcal{F}^t_{-}&=&z^{b\al_1}(1-z)^{1-b(\al_3-b)}\times\nonumber\\
&&_2F_1(b(\al_1-\al_3-\al_4+b/2)+1,b(\al_1-\al_3+\al_4-b/2);2-b(2\al_3-b);1-z) \, .\nonumber
\end{eqnarray}
The computation of the monodromy of these quantities is then standard.  Let
us introduce
$$
\fus{\al_1}{-b/2}{\al_3}{\al_4}{\al_1-sb/2,}{\al_3-s'b/2} \equiv
F^{L}_{s,s'},\quad s,s'=\pm.
$$
Then the entries of this $2\times 2$ matrix are:
\begin{eqnarray}
F_{++}&=&\frac{\Gamma(b(2\al_1-b))\Gamma(b(b-2\al_3)+1)}
{\Gamma(b(\al_1-\al_3-\al_4+b/2)+1)\Gamma(b(\al_1-\al_3+\al_4-b/2))} \nonumber \\
F_{+-}&=& \frac{\Gamma(b(2\al_1-b))\Gamma(b(2\al_3-b)-1)}
{\Gamma(b(\al_1+\al_3+\al_4-3b/2)-1)\Gamma(b(\al_1+\al_3-\al_4-b/2))} \nonumber \\
F_{-+}&=&\frac{\Gamma(2-b(2\al_1-b))\Gamma(b(b-2\al_3)+1)}
{\Gamma(2-b(\al_1+\al_3+\al_4-3b/2))\Gamma(1-b(\al_1+\al_3-\al_4-b/2))} \nonumber \\
F_{--}&=& \frac{\Gamma(2-b(2\al_1-b))\Gamma(b(2\al_3-b)-1)}
{\Gamma(b(-\al_1+\al_3+\al_4-b/2))\Gamma(b(-\al_1+\al_3-\al_4+b/2)+1)} \nonumber \\
\label{f1/2} \nonumber
\end{eqnarray}
The dual case with $\al_2=-b^{-1}/2$ is obtained by substituting
$b$ by $b^{-1}$.
\end{itemize}

\subsubsection{Construction of the fusion matrix: strategy \cite{PT1}}
We now make the following change of normalization that preserves
the Moore-Seiberg equations:
\begin{eqnarray}
\fus{\al_1}{\al_2}{\al_3}{\al_4}{\al_{21}}{\al_{32}} =
\frac{N(\al_4,\al_3,\al_{21})
N(\al_{21},\al_2,\al_1)}{N(\al_4,\al_{32},\al_1)
N(\al_{32},\al_3,\al_2)}
\gfus{\al_1}{\al_2}{\al_3}{\al_4}{\al_{21}}{\al_{32}}
\label{passage de Fus a Gus}
\end{eqnarray}
where the function $N(\al_3,\al_2,\al_1)$ reads
\begin{eqnarray}
\lefteqn{N(\al_3,\al_2,\al_1) =}\nonumber \\
&&
\frac{\Ga(2\al_1)\Ga(2\al_2)\Ga(2Q-2\al_3)}{\Ga(2Q-\al_1-\al_2-\al_3)
\Ga(\al_1+\al_2-\al_3)\Ga(\al_1+\al_3-\al_2)\Ga(\al_2+\al_3-\al_1)}\, .
\nonumber \label{normalisation}
\end{eqnarray}
Now the crossing symmetry condition takes the form of an
orthogonality relation for
$\gfus{\al_1}{\al_2}{\al_3}{\al_4}{\al_{21}}{\al_{32}}~:$
\begin{equation}
\int_{\Sp}d\al_{21}M_{b}(\al_{21})\gfus{\al_1}{\al_2}{\al_3}{\al_4}{\al_{21}}{\al_{32}}
\left(\gfus{\al_1}{\al_2}{\al_3}{\al_4}{\al_{21}}{\beta_{32}}\right)^{*}
= M_{b}(\al_{32})\delta(\al_{32}-\beta_{32}) \label{ortho
G}\nonumber
\end{equation}
where
\begin{eqnarray}
M_b(\al)=-4\text{sin}\pi b(2\al-Q)\text{sin}\pi
b^{-1}(2\al-Q).\label{mesureSb}
\end{eqnarray}
If $\al_2=-b/2$, the matrix $G_{s,s'},\; s,s'=\pm$ becomes
\begin{eqnarray}
G_{s,s'}=\left(
\begin{array}
[c]{cc}%
\frac{[\al_4+\al_3-\al_1-\frac{b}{2}]_{b}}{[2\al_3-b]_{b}} & \frac{[\al_4+\al_3+\al_1-\frac{3b}{2}]_{b}}{[2\al_3-b]_{b}} \\
\frac{[\al_3+\al_1-\al_4-\frac{b}{2}]_{b}}{[2\al_3-b]_{b}} & -\frac{[\al_4+\al_1-\al_3-\frac{b}{2}]_{b}}{[2\al_3-b]_{b}} \\%
\end{array}
\right) \label{g1/2}
\end{eqnarray}
where $[x]_{b}=\frac{\text{sin}(\pi bx)}{\text{sin}\pi b^{2}}$.
The entries of the matrix are thus purely trigonometrical.
\vspace{0.5cm}

To construct the fusion matrix explicitly, we propose to do the
following:
\begin{itemize}
  \item
  Choose an appropriate quantum group: $\uq$ with deformation parameter
  $q=e^{i\pi b^2}, \, |q|=1$, seems to be a reasonable candidate to describe
  the internal structure of Liouville field theory.
  \item
  Construct the Clebsch-Gordan coefficients, then the $6j$ symbol (or
  Racah-Wigner coefficients), for the continuous unitary
  representations $\mathcal{P}_{\al}, \; \al=\frac{Q}{2}+iP$
  of $\uq$.
  \item
  Find a suitable normalization for the Clebsch-Gordan
  coefficients such that the $6j$ symbol constructed
  coincides with the function
  $\gfus{\al_1}{\al_2}{\al_3}{\al_4}{\al_{21}}{\al_{32}}$ (this will be
   done by comparing the found expression for the $6j$ symbol
   for the special case $\al_2=-b/2$ with (\ref{g1/2})).
\end{itemize}
We would like to emphasize that:
\begin{itemize}
  \item
  Such a construction by means of representation theory methods will {\it
ensure}
  the validity of the Moore-Seiberg equations for our fusion
  matrix.
  \item
  {\bf The orthogonality condition for the
  Racah-Wigner coefficients is equivalent to proving crossing
  symmetry in Liouville field theory}.
\end{itemize}

\section{Clebsch-Gordan coefficients and Racah-Wigner symbol for a
continuous series of representations of $\uq$ \cite{PT1,PT2}}

\subsection{A class of representations of $\uq$}
$\mathcal{U}_q(\mathfrak{sl}(2))$ is a Hopf algebra with
generators $E,\; F,\; K,\; K^{-1}$ that satisfy the
$q$-commutation relations
$$
\begin{array}{lll}
KE=qEK, & KF=q^{-1}FK, & [E,F]=-\frac{K^{2}-K^{-2}}{q-q^{-1}}\, .
\end{array}
\label{relations de commutation}
$$
The coproduct is
$$
\begin{array}{lll}
\Delta(K)=K\otimes K,   & \Delta(E)=E\otimes K+ K^{-1}\otimes E, &
\Delta(F)=F\otimes K + K^{-1}\otimes F.
\end{array}
$$
The center of $\mathcal{U}_q(\mathfrak{sl}(2))$ is generated by
the $b$-Casimir
\begin{equation}
C= FE -\frac{qK^{2}+q^{-1}K^{-2}-2}{(q-q^{-1})^{2}}\, .\nonumber
\end{equation}
The real form $\uq, \,q=e^{i\pi b^2},\, |q|=1$, has the following star
structure for its unitary representations
$$
\begin{array}{lll}
K^{*}=K, & F^{*}=F, & E^{*}=E.
\end{array}
$$
We realize the set of unitary continuous representations
$\mathcal{P}_{\al},\; \al=\frac{Q}{2}+iP$ on the Hilbert space
$L^{2}(\mathbb{R})$ in terms of the Weyl algebra generated by
$\mathrm{U}=e^{2\pi bx}$ and $\mathrm{V}=e^{-\frac{b}{2}p}$ where
$[x,p]=i$:
\begin{eqnarray}
E&=&\mathrm{U}^{+1}\frac{e^{i\pi b(Q-\al)}\mathrm{V}-e^{-i\pi b(Q-\al)}\mathrm{V}^{-1}}{e^{i\pi b^{2}}-e^{-i \pi b^{2}}} \; ,\nonumber\\
F&=&\mathrm{U}^{-1}\frac{e^{-i\pi b(Q-\al)}\mathrm{V}-e^{i\pi b(Q-\al)}\mathrm{V}^{-1}}{e^{i\pi b^{2}}-e^{-i \pi b^{2}}} \, ,\nonumber\\
K&=&\mathrm{V} \, .\nonumber  \label{def des generateurs}
\end{eqnarray}
where $\mathrm{U}$ is the operator of multiplication by $e^{2\pi b
x}$ and $\mathrm{V}$ is the shift operator acting on a function
$f(x)$ of the real variable $x$ as
$\mathrm{V}f(x)=f(x+i\frac{b}{2})$. Note that we are dealing with unbounded operators, as the quantum group is non compact.\\
\underline{Remark}:\\
In the limit where $b \to 0,\; \al \sim -bj,\; z=e^{2\pi b x} $,
one has
\begin{eqnarray}
E \sim z^2\partial_z+ (j+1)z,\quad F \sim
\partial_z-(j+1)z^{-1},\quad \frac{K^2-K^{-2}}{q-q^{-1}} \sim
2\partial_z. \nonumber
\end{eqnarray}
\vspace{0.2cm}

The representation will be realized on the space
$\mathcal{P}_{\al}$ of entire analytic functions $f(x)$ that have
a Fourier transform $f(\omega)$ which is meromorphic in
$\mathbb{C}$ with possible poles at
\begin{eqnarray}
\omega &=& i(\al -Q-nb-mb^{-1}),\nonumber \\
\omega &=& i(Q-\al +nb+mb^{-1}).\nonumber
\end{eqnarray}
They are unitarily equivalent to a subset of the integrable
representations of $\uq$ that appear in the classification of
\cite{schmudgen2}. In particular, the generators have self-adjoint
extensions in $L^{2}(\mathbb{R})$. The $b$-Casimir acts as a
scalar in this representation
\begin{equation}
C= FE -\frac{qK^{2}+q^{-1}K^{-2}-2}{(q-q^{-1})^{2}}\equiv
[\al-\frac{Q}{2}]_{b}^{2},\quad [x]_b \equiv \frac{\sin \pi bx}{\sin
\pi b^2}. \nonumber
\end{equation}
\underline{Remark}:\\
The representations we consider here have the remarkable property
that if one introduces $\tilde{E},\tilde{F},\tilde{K}$ by
replacing $b \to 1/b$ in the expressions for $E,F,K$ given above,
one obtains a representation of
$\mathcal{U}_{\tilde{q}}(\mathfrak{sl}(2,\mathbb{R}))$,
$\tilde{q}=e^{i\pi b^{-2}}$ on the same space $\mathcal{P}_{\al}$.
On this space, the generators of the two dual quantum groups
commute. This autoduality of our representations
$\mathcal{P}_{\al}$ is related to the fact that they {\it do not}
have a classical limit when $b \to 0$. This is after all not so
surprising, because autoduality in Liouville field theory is a
characteristic of the quantum theory only. Lastly, we note that
representations $\mathcal{P}_{\al}$ and $\mathcal{P}_{Q-\al}$ are
equivalent.

\subsection{Clebsch-Gordan decomposition of tensor products}
The Clebsch-Gordan map $C(\al_3|\al_2,\al_1):\p{2} \otimes \p{1}
\rightarrow \p{3}$ can be explicitly represented by an integral
transform
$$
f(x_2,x_1) \to F[f](\al_3,x_3)\equiv
\int\limits_{\mathbb{R}}dx_2dx_1
\clebsch{\al_3}{\al_2}{\al_1}{x_3}{x_2}{x_1}f(x_2,x_1)  \nonumber
$$
The distributional kernel $\clebsch{\al_3}{\al_2}{\al_1}{x_3}{x_2}{x_1}$ is the
Clebsh-Gordan coefficients ($3j$ symbol). It is an invariant with
respect to the action of the quantum group and satisfies
\begin{eqnarray}
 K_3\clebsch{\al_3}{\al_2}{\al_1}{x_3}{x_2}{x_1}&=&
 \Delta_{21}^{t}(K)\clebsch{\al_3}{\al_2}{\al_1}{x_3}{x_2}{x_1}\nonumber\\
 E_3\clebsch{\al_3}{\al_2}{\al_1}{x_3}{x_2}{x_1}&=&
 \Delta_{21}^{t}(E)\clebsch{\al_3}{\al_2}{\al_1}{x_3}{x_2}{x_1}\nonumber\\
 F_3\clebsch{\al_3}{\al_2}{\al_1}{x_3}{x_2}{x_1}&=&
 \Delta_{21}^{t}(F)\clebsch{\al_3}{\al_2}{\al_1}{x_3}{x_2}{x_1}\nonumber
\end{eqnarray}
where $\Delta^{t}$ means transposed action. These equations can be
solved with the following result \cite{PT1}:
\begin{eqnarray}
\lefteqn{\clebsch{\al_3}{\al_2}{\al_1}{x_3}{x_2}{x_1}=e^{-2\pi
(Q-\al_3)x_3
-2\pi(\al_1+\al_3-Q)x_2+2\pi\al_1 x_1} M(\al_3,\al_2,\al_1)} \nonumber \\
& & \lim_{\epsilon \rightarrow 0^{+}}\frac{G_b(-i(x_1-x_3)+\frac{1}{2}(2\al_2+\al_3-\al_1))}{G_b(-i(x_1-x_3)+\frac{1}{2}(\al_1-\al_3)+Q-\epsilon)}\frac{G_b(-i(x_1-x_2)+\frac{1}{2}(-\al_1-\al_2)+Q)}{G_b(-i(x_1-x_2)+\frac{1}{2}(\al_1+\al_2+2\al_3)-\epsilon)}. \nonumber \\
&
&\frac{G_b(-i(x_2-x_3)-\frac{1}{2}(\al_2-\al_3)+\epsilon)}{G_b(-i(x_2-x_3)-\frac{1}{2}(2\al_1-\al_2+\al_3)+Q)}\; .
 \label{expression3j}\nonumber
\end{eqnarray}
We introduced the function $G_b(x)=e^{\frac{i\pi}{2}x(x-Q)}S_b(x)$
(see Appendix B for the definition of $S_b$) that satisfies the
functional relation $G_b(x+b)=\left(1-e^{2i\pi bx}\right)G_b(x)$,
as well as the dual equation with $b$ replaced by $1/b$. The
coefficient $M(\al_3,\al_2,\al_1)$ is a constant that normalizes
the $3j$ symbol. We choose, for reasons that will be explained
later,
\begin{equation}
M(\al_3,\al_2,\al_1)=e^{i\pi
(\al_{1}^2+\al_1\al_3-Q\al_1)}S_b(\al_1+\al_2-\al_{3})\, .\label{M}
\end{equation}
It follows from the construction of a common spectral
decomposition \cite{PT2} for the Casimir operator $C_{21}$ and the
third generator $K_{21}$ that the Clebsch-Gordan coefficients
satisfy the orthogonality and completeness relations:
\begin{eqnarray}
\int_{\mathbb{R}}dx_1dx_2\clebsch{\al_{3}}{\al_{2}}{\al_{1}}{x_{3}}{x_{2}}{x_{1}}^{*}
\clebsch{\beta_{3}}{\al_{2}}{\al_{1}}{y_{3}}{x_{2}}{x_{1}}=|S_b(2\al_3)|^{-2}\delta(\al_3-\beta_3)\delta(x_3-y_3)
 \nonumber
\label{orthoclebsch}
\end{eqnarray}
Note that $|S_b(2\al)|^{2} \equiv M_{b}(\al)$ (see equation
(\ref{mesureSb})).
\begin{eqnarray}
\int_{\Sp}d\al_{3}|S_b(2\al_3)|^{2}\int_{\mathbb{R}}dx_3\clebsch{\al_{3}}{\al_{2}}{\al_{1}}{x_{3}}{x_{2}}{x_{1}}^{*}
\clebsch{\al_{3}}{\al_{2}}{\al_{1}}{x_{3}}{y_{2}}{y_{1}}=\delta(x_2-y_2)\delta(x_1-y_1)
\nonumber  \label{completude des clebsch}
\end{eqnarray}
From the completeness relation it follows that our {\bf autodual
representations are closed under tensor product}, which is {\it a
priori} non trivial if there are other unitary representations of
$\uq$.

\subsection{Racah-Wigner coefficients for $\uq$}

It is possible to construct two canonical bases of  $\p{3} \otimes
\p{2} \otimes \p{1}$. The decomposition of this representation into irreducibles can be done in two
 different ways, by iteration of the
Clebsch-Gordan mapping
\begin{itemize}
\item
either by using $(id\otimes\Delta)\circ \Delta$
\item
or $(\Delta \otimes id)\circ \Delta$
\end{itemize}
The expression for the first base in the $s$-channel is given by
\begin{eqnarray}
\phis{\al_1}{\al_2}{\al_3}{\al_4}{\al_{21}}=\int_{\mathbb{R}}dx_{21}
\clebsch{\al_4}{\al_3}{\al_{21}}{x_{4}}{x_{3}}{x_{21}}\clebsch{\al_{21}}{\al_{2}}{\al_{1}}{x_{21}}{x_{2}}{x_{1}}
\label{conv clebsch s}
\end{eqnarray}
and similarly for the $t$-channel base
\begin{eqnarray}
\phit{\al_1}{\al_2}{\al_3}{\al_4}{\al_{32}}=\int_{\mathbb{R}}dx_{32}\clebsch{\al_4}{\al_{32}}{\al_{1}}{x_{4}}{x_{32}}{x_{1}}\clebsch{\al_{32}}{\al_{3}}{\al_{2}}{x_{32}}{x_{3}}{x_{2}}
\label{conv clebsch t}
\end{eqnarray}
From an argument of completeness and orthogonality \cite{PT2} it follows that the bases $\Phi^{s}$
and  $\Phi^{t}$ are related by a transformation
\begin{eqnarray}
\phis{\al_1}{\al_2}{\al_3}{\al_4}{\al_{21}}=\int_{\Sp}d\al_{32}\my6j{\al_1}{\al_2}{\al_3}{\al_4}{\al_{21}}{\al_{32}}\phit{\al_1}{\al_2}{\al_3}{\al_4}{\al_{32}}
\label{def des 6j}\nonumber
\end{eqnarray}
defining thus the Racah-Wigner coefficients. Their construction
implies the orthogonality condition which is equivalent to proving
crossing-symmetry in Liouville field theory
\begin{eqnarray}
\int_{\Sp}|S_{b}(\al_{21})|^{2}\my6j{\al_1}{\al_2}{\al_3}{\al_4}{\al_{21}}{\al_{32}}\my6j{\al_1}{\al_2}{\al_3}{\al_4}{\al_{21}}{\beta_{32}}^{*}
= |S_{b}(\al_{32})|^{2}\delta(\al_{32}-\beta_{32})\nonumber
\label{orthogonalite des 6j}
\end{eqnarray}

\subsection{Explicit computation of the Racah-Wigner coefficients}
It seems {\it a priori} difficult to compute the integrals
(\ref{conv clebsch s}) and (\ref{conv clebsch t}), as
 $\Phi^{s},\Phi^{t}$ are functions of the four variables $x_4,x_3,x_2,x_1\;$;
 however, one simplifies the problem in the limit where $x_4 \rightarrow +\infty$ and $x_2 \rightarrow -\infty$
 by making use of the asymptotics of the $S_b$ function
\begin{eqnarray}
S_b(x) \sim e^{\mp \frac{i\pi}{2}(x^2-xQ)},\quad  \mathrm{Im}x
\rightarrow \pm \infty. \nonumber
\end{eqnarray}
The result is then expressed in terms of $x_1$ and
 $x_3$. It turns out that it is convenient to set $x_3=\frac{i}{2}(\al_2+\al_4)$. Let $x \equiv x_1$.\\
We obtain in the $s$-channel
\begin{eqnarray}
\psis{\al_1}{\al_2}{\al_3}{\al_4}{\al_{21}}=\norms{\al_1}{\al_2}{\al_3}{\al_4}{\al_{21}}
\thetas{\al_1}{\al_2}{\al_3}{\al_4}{\al_{21}}\nonumber
\label{ecriture de psis}
\end{eqnarray}
with
\begin{eqnarray}
\nonumber \\
\norms{\al_1}{\al_2}{\al_3}{\al_4}{\al_{21}} \makebox[1.4 em]{}
 &=& \frac{M(\al_{21},\al_2,\al_1)M(\al_4,\al_3,\al_{21})}{S_b(2\al_{21})} S_b(\al_{21}+\al_2-\al_1)
 S_b(\al_{21}+\al_1-\al_2) \nonumber \\
&& \times e^{-i\pi [\al_4(\al_2+\al_4+\al_3+\al_{21}-Q)+\al_{21}(\al_1+\al_{21}-Q)+\al_3\al_2]}  \nonumber \\
\thetas{\al_1}{\al_2}{\al_3}{\al_4}{\al_{21}}
 &=& e^{2\pi(\al_{21}-\al_1-\al_2)x}  \nonumber \\
&& \mbox{\em \em \em}\times
F_{b}(\al_{21}+\al_1-\al_2,\al_{21}+\al_3+\al_4-Q;2\al_{21};-ix)
\nonumber
\end{eqnarray}
where the function $F_b(a,b;c;ix)$ is the $b$-deformed
hypergeometric function in the Euler representation defined in the
Appendix C.

Similarly, we have in the $t$-channel
\begin{eqnarray}
\psit{\al_1}{\al_2}{\al_3}{\al_4}{\al_{32}}=\normt{\al_1}{\al_2}{\al_3}{\al_4}{\al_{32}}
\thetat{\al_1}{\al_2}{\al_3}{\al_4}{\al_{32}}\nonumber
\label{ecriture de psit}
\end{eqnarray}
with
\begin{eqnarray}
\nonumber \\
\normt{\al_1}{\al_2}{\al_3}{\al_4}{\al_{32}}  \makebox[1.4 em]{}
&=& \frac{M(\al_{32},\al_3,\al_{2})M(\al_{4},\al_{32},\al_1)}{S_b(2\al_{32})}  S_b(\al_{32}+\al_3-\al_2) \nonumber \\
&& S_b(\al_{32}+\al_2-\al_3) e^{-i\pi[\al_4(\al_1+\al_4+\al_2+\al_3-Q) +\al_2(\al_{32}+\al_3+\al_2-Q)]}\nonumber \\
\thetat{\al_1}{\al_2}{\al_3}{\al_4}{\al_{32}}
&=& e^{-2\pi(\al_{32}+\al_1+\al_4-Q)x} \times \nonumber \\
&& F_{b}(\al_{32}+\al_1+\al_4-Q,\al_{32}+\al_3-\al_2;2\al_{32};ix)
\nonumber
\end{eqnarray}

Now we use the normalization
\begin{eqnarray}
\int_{\mathbb{R}}dxe^{2\pi
Qx}\left(\thetas{\al_1}{\al_2}{\al_3}{\al_4}{\al_{21}'}\right)^{*}
\thetas{\al_1}{\al_2}{\al_3}{\al_4}{\al_{21}}=\delta(\al_{21}-\al_{21}')\, ,
\nonumber
\end{eqnarray}
and the fact that $\left\lbrace\Theta_{\al_{21}}, \al_{21}\in
\Sp\right\rbrace$ and $\left\lbrace\Theta_{\al_{32}}, \al_{32}\in
\Sp\right\rbrace$ form complete sets of generalized
eigenfunctions for the operators $C_{21}$ and $C_{32}$ respectively
\cite{PT2}
\begin{eqnarray}
&& \mathcal{C}_{21}=[\delta_x+\al_1+\al_2-\frac{Q}{2}]_{b}^{2}-
e^{2\pi bx}[\delta_x+\al_1+\al_2+\al_3+\al_4-Q]_{b}[\delta_x+2\al_1]_{b} \nonumber \\
&& \mathcal{C}_{32}=[\delta_x+\al_1+\al_4-\frac{Q}{2}]_{b}^{2}-
e^{-2\pi bx}[\delta_x+\al_1+\al_2-\al_3+\al_4-Q]_{b}[\delta_x]_{b}
\nonumber
\end{eqnarray}
where $\delta_{x}=\frac{1}{2\pi}\partial_{x},\;
[x]_{b}=\frac{\text{sin}(\pi bx)}{\text{sin}\pi b^{2}}$. \\
We get
the following expression for the Racah-Wigner symbol:
\begin{eqnarray}
\lefteqn{\my6j{\al_1}{\al_2}{\al_3}{\al_4}{\al_{21}}{\al_{32}}=
\frac{\norms{\al_1}{\al_2}{\al_3}{\al_4}{\al_{21}}}{\normt{\al_1}{\al_2}{\al_3}{\al_4}{\al_{32}}}\quad\times}\nonumber \\
&& \int_{\mathbb{R}}e^{2\pi
Qx}\left(\thetat{\al_1}{\al_2}{\al_3}{\al_4}{\al_{32}}\right)^{*}
\thetas{\al_1}{\al_2}{\al_3}{\al_4}{\al_{21}}.\nonumber
\label{6j
en terme de 2F1}
\end{eqnarray}
Using now the Barnes representation for the $b$-deformed hypergeometric
function $_2F_1$ presented in the Appendix C, the Racah-Wigner
coefficients can be expressed explicitly in terms of a
$b$-deformed hypergeometric function $_4F_3$ in the Barnes
representation:

\begin{eqnarray}
\lefteqn{\my6j{\al_1}{\al_2}{\al_3}{\al_4}{\al_{21}}{\al_{32}}=} \nonumber \\
&& \frac{S_b(\al_2+\al_{21}-\al_{1})S_b(\al_2+\al_{1}-\al_{21})S_b(\al_{21}+\al_3-\al_{4})S_b(\al_{32}+\al_{1}+\al_{4}-Q)}{S_b(\al_2+\al_{32}-\al_{3})S_b(\al_{3}+\al_2-\al_{32})S_b(\al_{21}+\al_3+\al_{4}-Q)S_b(\al_{32}+\al_{1}-\al_{4})}. \nonumber \\
&&|S_b(2\al_{32})|^{2}\int\limits_{-i\infty}^{i\infty}ds \;\;
\frac{S_b(U_1+s)S_b(U_2+s)S_b(U_3+s)S_b(U_4+s)}
{S_b(V_1+s)S_b(V_2+s)S_b(V_3+s)S_b(Q+s)} \nonumber
\label{formule
de F}
\end{eqnarray}
where:
$$
\begin{array}{ll}
 U_1 = \al_{21}+\al_1-\al_2                &  V_1 = Q+\al_{21}-\al_{32}-\al_{2}+\al_{4} \\
 U_2 = Q+\al_{21}-\al_2-\al_1              &  V_2 = \al_{21}+\al_{32}+\al_{4}-\al_2 \\
 U_3 = \al_{21}+\al_{3}+\al_{4}-Q          &  V_3 = 2\al_{21} \\
 U_4 = \al_{21}-\al_{3}+\al_4              \\
\end{array}
$$
Now we set $\al_2=-b/2$: the fusion rules are
$$\left\lbrace
\begin{array}{l}
\al_{21} = \al_{1}-s\frac{b}{2}\, , \\
\al_{32} = \al_{3} -s'\frac{b}{2}\, ,\quad s,s'=\pm \; .
\end{array}\right.$$
In this case some poles of the integrand will cross the imaginary
axis. It is not difficult to compute the residues, and one finds
\begin{eqnarray}
\my6j{\al_1}{-b/2}{\al_3}{\al_4}{\al_1-sb/2}{\al_3-s'b/2}= \left(
\begin{array}
[c]{cc}%
\frac{[\al_4+\al_3-\al_1-\frac{b}{2}]_{b}}{[2\al_3-b]_{b}} & \frac{[\al_4+\al_3+\al_1-\frac{3b}{2}]_{b}}{[2\al_3-b]_{b}} \\
\frac{[\al_3+\al_1-\al_4-\frac{b}{2}]_{b}}{[2\al_3-b]_{b}} & -\frac{[\al_4+\al_1-\al_3-\frac{b}{2}]_{b}}{[2\al_3-b]_{b}} \\%
\end{array}
\right)\nonumber
\end{eqnarray}
The normalization $M(\al_3,\al_2,\al_1)$ introduced in equation (\ref{M}) has been chosen
such that the Racah-Wigner coefficient equals $G_{s,s'}$ for
$\al_2=-b/2$. It follows from a uniqueness argument for the fusion
matrix (see \cite{PT1} and Appendix C of \cite{PT3}) that these
two objects are equal for any values of $\al_i$.

\subsection{Expression of the fusion matrix} It is now shown that
$\gfus{\al_1}{\al_2}{\al_3}{\al_4}{\al_{21}}{\al_{32}}=\my6j{\al_1}{\al_2}{\al_3}{\al_4}{\al_{21}}{\al_{32}}$,
so we deduce thanks to equation (\ref{passage de Fus a Gus}) the
expression for the fusion matrix \cite{PT1}:

\begin{eqnarray}
\lefteqn{\fus{\al_1}{\al_2}{\al_3}{\al_4}{\al_{21}}{\al_{32}}=} \nonumber \\
&&
\frac{\Ga(2Q-\al_3-\al_2-\al_{32})\Ga(\al_3+\al_{32}-\al_2)\Ga(Q-\al_2-\al_{32}+\al_3)\Ga(Q-\al_3-\al_2+\al_{32})}{\Ga(2Q-\al_1-\al_2-\al_{21})\Ga(\al_1+\al_{21}-\al_2)\Ga(Q-\al_2-\al_{21}+\al_1)\Ga(Q-\al_2-\al_1+\al_{21})} \nonumber \\
&&
\frac{\Ga(Q-\al_{32}-\al_1+\al_4)\Ga(\al_{32}+\al_{1}+\al_{4}-Q)\Ga(\al_1+\al_4-\al_{32})\Ga(\al_4+\al_{32}-\al_1)}{\Ga(Q-\al_{21}-\al_{3}+\al_4)\Ga(\al_{21}+\al_3+\al_4-Q)\Ga(\al_3+\al_4-\al_{21})\Ga(\al_{21}+\al_4-\al_3)} \nonumber \\
&&\frac{\Ga(2Q-2\al_{21})\Ga(2\al_{21})}{\Ga(Q-2\al_{32})\Ga(2\al_{32}-Q)}
\frac{1}{i}\int\limits_{-i\infty}^{i\infty}ds \;\;
\frac{S_b(U_1+s)S_b(U_2+s)S_b(U_3+s)S_b(U_4+s)}
{S_b(V_1+s)S_b(V_2+s)S_b(V_3+s)S_b(Q+s)} \nonumber
\label{formule
de F}
\end{eqnarray}
where:
$$
\begin{array}{ll}
 U_1 = \al_{21}+\al_1-\al_2                &  V_1 = Q+\al_{21}-\al_{32}-\al_{2}+\al_{4} \\
 U_2 = Q+\al_{21}-\al_2-\al_1              &  V_2 = \al_{21}+\al_{32}+\al_{4}-\al_2 \\
 U_3 = \al_{21}+\al_{3}+\al_{4}-Q          &  V_3 = 2\al_{21} \\
 U_4 = \al_{21}-\al_{3}+\al_4 \\              \\
\end{array}
$$
One can check explicitly that the fusion matrix is indeed
invariant with respect to the exchange of rows and columns, and
that it depends on conformal weights only. Some of these
properties are trivial to check, others requires the use of
transformation properties of $b$-deformed hypergeometric functions
like the one presented in the Appendix C.

\section{Application: Boundary three point function \cite{PT3}}
We now consider Liouville theory on a simply connected domain
$\Gamma$ with a non trivial boundary $\partial \Gamma$. $\Gamma$
can be either the unit disk, the upper half-plane, or the infinite
strip. A conformally invariant boundary condition in LFT can be
introduced through the following boundary interaction
\begin{equation}
A_{\mathrm{bound}}=A_{\mathrm{bulk}}+\int\limits_{\partial\Gamma}\left(
\frac{QK}{2\pi}\phi+\mu_{B}e^{b\phi}\right)
g^{1/4}dx \, ,\label{gbound} \nonumber
\end{equation}
where the integration in $x$ is along the boundary while $K$ is
the curvature of the boundary in the background geometry $g$. We
will consider the geometry of the Upper Half Plane (flat metric
background)
\begin{equation}
A_{\mathrm{bound}}=\int\limits_{\mathrm{UHP}}\left(  \frac{1}{4\pi}(\partial_{a}%
\phi)^{2}+\mu e^{2b\phi(z,\bar{z})}d^{2}z\right)
+\mu_{B}\int\limits_{\mathbb{R}}e^{b\phi(x)}  dx\label{bound}
\nonumber
\end{equation}
with the boundary condition at $|z|\rightarrow \infty$
\begin{equation}
\phi(z,\bar{z})=-Q\log(z\bar{z})+O(1),\label{bcharge} \nonumber
\end{equation}
and the Neumann condition for the Liouville field $\phi$ on the real
axis. We call $\mu_{B}$ the boundary cosmological constant, by
analogy with the bulk case \footnote{A different set of boundary conditions is
 considered in the work \cite{ZZ2}, following early works of \cite{DJ,DJ2,DFJ}.}.
  There is actually a one-parameter family of conformally invariant boundary conditions characterized
by different values of the boundary cosmological constant $\mu_B$.
We shall denote the boundary operators
$B_{\beta}^{\sigma_{2}\sigma_{1}}(x)$ (classically they correspond
to the boundary value of $e^{\beta/2b\varphi}\;$, where $\varphi$
is the classical Liouville field: $\varphi=2b\phi$). The boundary
operators have conformal weight $\Delta_{\beta}=\beta(Q-\beta)$,
and are labelled by two left and right boundary conditions
$\sigma_{1}$ and $\sigma_{2}$ related to $\mu_{B_1}$ and
$\mu_{B_2}$ by the relation \cite{FZZ}
\begin{eqnarray}
\text{cos}\left(2\pi
b(\sigma-\frac{Q}{2})\right)=\frac{\mu_{B}}{\sqrt{\mu}}\sqrt{\text{sin}(\pi
b^{2})}. \label{relation mu-sigma}
\end{eqnarray}
Let us note for real values of $\mu_{B}$ the two following regimes
for the parameter $\sigma$:
\begin{enumerate}
\item
if $\frac{\mu_{B}}{\sqrt{\mu}}\sqrt{\text{sin}(\pi b^{2})}>1$,
then $\sigma$ is of the form $\sigma= Q/2 + iP,$
\item
if $\frac{\mu_{B}}{\sqrt{\mu}}\sqrt{\text{sin}(\pi b^{2})}<1$,
then $\sigma$ is real.
\end{enumerate}
Anticipating that all relevant objects will be found to possess
meromorphic continuation with respect to the boundary parameter
$\sigma$, we shall discuss only the first regime explicitly in the
following. The Hilbert space was found \cite{tesch1} to decompose
into irreducible
 representations of the Virasoro algebra:
\begin{eqnarray}
\mathcal{H}^B=\int_{\Sp}^{\oplus}\mathcal{V}_{\beta}.\nonumber
\end{eqnarray}
Contrary to the pure bulk case situation where the cosmological
constant enters only as a scale parameter, in the boundary case,
the observables depend on the scale invariant ration
$\mu/\mu_B^2$: for example, a correlation function with the bulk
operators $V_{\alpha_1} V_{\alpha_2}\dots V_{\alpha_n}$ and the
boundary operators
$B_{\beta_1}^{\sigma_{1}\sigma_{2}}B_{\beta_2}^{\sigma_{2}\sigma_{3}}\dots
B_{\beta_m}^{\sigma_{m}\sigma_{1}}$ scales as follow
\begin{eqnarray}
\mathcal{G}(\alpha_1,\dots\alpha_n,\beta_1\dots\beta_m)\sim
\mu^{(Q-2\sum_i \alpha_i-\sum_j \beta_j)/2b}
F\left(\frac{\mu^2_{B_1}}{\mu},\frac{\mu^2_{B_2}}{\mu},\dots,
\frac{\mu^2_{B_m}}{\mu}\right)\; ,\nonumber
\end{eqnarray}
where $F$ is some scaling function. The observables are autodual
provided the dual cosmological constant $\mu$ is related to
$\tilde{\mu}$ as in (\ref{mu}), and the dual boundary cosmological
constant is defined as follows
$$
\cos\left(\frac{2\pi}
{b}(\sigma-\frac{Q}{2})\right)=\frac{\tilde{\mu}_{B}}{\sqrt{\tilde{\mu}}}\sqrt{\sin\frac{\pi}
{b^2}}\;.
$$
 In order to characterize LFT on the upper half plane,
one needs to know additional structure constants beside the bulk
three point
  function $C(\al_1,\al_2,\al_3)$.
\begin{enumerate}
\item bulk one point function \cite{FZZ,T3}
\begin{equation}
\left\langle V_{\alpha}(z,\bar{z})\right\rangle
=\frac{U(\alpha|\mu_{B})}{\left| z-\bar{z}\right|
^{2\Delta_{\alpha}}} \label{onepoint} \nonumber
\end{equation}
\item boundary two point function \cite{FZZ}
\begin{equation}
\left\langle
B_{\beta_1}^{\sigma_{1}\sigma_{2}}(x)B_{\beta_2}^{\sigma_{2}\sigma_{1}}(0)\right\rangle
=\frac{\delta(\beta_2+\beta_1-Q)+S(\beta_1,\sigma_{2},\sigma_{1})\delta(\beta_2-\beta_1)}{\left|
x\right|^{2\Delta_{\beta_1}}} \nonumber
\end{equation}
We give here the explicit expression for the boundary reflection
amplitude:
\begin{eqnarray}
\lefteqn{S(\beta,\sigma_{2},\sigma_{1})
 =(\pi\mu\gamma(b^{2})b^{2-2b^{2}})^{\frac{1}{2b}(Q-2\beta)}}
\nonumber \\
& & \frac{\Gamma_{b}(2\beta-Q)}{\Gamma_{b}(Q-2\beta)}
\frac{S_b(\sigma_{2}+\sigma_{1}-\beta)S_b(2Q-\beta-\sigma_{1}-\sigma_{2})}
{S_b(\beta+\sigma_{2}-\sigma_{1})S_{b}(\beta+\sigma_{1}-\sigma_{2})}
\nonumber  \label{fonction 2 points boundary}
\end{eqnarray}
It satisfies the unitarity relation
$S(\beta,\sigma_{2},\sigma_{1})S(Q-\beta,\sigma_{2},\sigma_{1})=1$.

\item bulk-boundary two point function\cite{hosomichi}\footnote{The bulk one point function is a special case of the
bulk-boundary coefficient with $\beta=0$.}
\begin{equation}
\left\langle
V_{\alpha}(z,\bar{z})B_{\beta}^{\sigma\sigma}(x)\right\rangle
=\frac{R(\alpha,\beta |\mu_{B})}{\left|  z-\bar{z}\right|
^{2\Delta_{\alpha}-\Delta_{\beta}}\left| z-x\right|
^{2\Delta_{\beta}}} \label{bbound} \nonumber
\end{equation}
\item boundary three point function
\begin{eqnarray}
\left\langle
B_{Q-\beta_{3}}^{\sigma_{1}\sigma_{3}}(x_{3})B_{\beta_{2}}^{\sigma_{3}\sigma_{2}}(x_{2})B_{\beta_{1}}^{\sigma_{2}
\sigma_{1}}(x_{1})\right\rangle
=\frac{C_{\beta_{2}\beta_{1}}^{(\sigma_{3}\sigma_{2}\sigma_{1})\beta_{3}}}{\left|
x_{21}\right|  ^{\Delta_{1}+\Delta_{2}-\Delta_{3}}\left|
x_{32}\right|
^{\Delta_{2}+\Delta_{3}-\Delta_{1}}\left|  x_{31}\right|  ^{\Delta_{3}%
+\Delta_{1}-\Delta_{2}}} \nonumber
\end{eqnarray}
\end{enumerate}
We now proceed to the determination of the latter quantity.

\subsection{Associativity condition}
Let us consider a four point function of boundary operators. This
quantity can be equivalently written\footnote{One restricts
oneself to the case where $Re(\beta_i), Re(\sigma_i), i=1 \dots 4
$ are close enough to Q/2. In this case, $\beta_{21}$ is of the
form $Q/2+iP$. Meromorphic continuation is understood otherwise.}:
\begin{itemize}
\item
in the $s$-channel
\begin{eqnarray}
\lefteqn{\left\langle B_{\beta_{4}}^{\sigma_{1}\sigma_{4}}(x_4)B_{\beta_3}^{\sigma_{4}\sigma_{3}}%
(x_3)B_{\beta_2}^{\sigma_{3}\sigma_{2}}(x_2)B_{\beta_1}^{\sigma_{2}\sigma_{1}}(x_1)%
\right\rangle =} \nonumber \\
&&\int_{\mathbb{S}}d\beta_{21}C_{\beta_{3},\beta_{21}}^{(\sigma_{4}\sigma_{3}\sigma_{1})\beta_{4}}
C_{\beta_{2}\beta_{1}}^{(\sigma_{3}\sigma_{2}\sigma_{1})\beta_{21}}
\mathcal{F}^{s}(\Delta_{\beta_i},\Delta_{\beta_{21}},x_i)
\nonumber  \label{4pointsBs}
\end{eqnarray}
\item
in the $t$-channel
\begin{eqnarray}
\lefteqn{\left\langle B_{\beta_{4}}^{\sigma_{1}\sigma_{4}}(x_4)B_{\beta_3}^{\sigma_{4}\sigma_{3}}%
(x_3)B_{\beta_2}^{\sigma_{3}\sigma_{2}}(x_2)B_{\beta_1}^{\sigma_{2}\sigma_{1}}(x_1)
\right\rangle =} \nonumber \\
&&
\int_{\mathbb{S}}d\beta_{32}C_{\beta_{32},\beta_{1}}^{(\sigma_{4}\sigma_{2}\sigma_{1})\beta_{4}}C_{\beta_{3}\beta_{2}}^{(\sigma_{4}\sigma_{3}\sigma_{2})\beta_{32}}
\mathcal{F}^{t}(\Delta_{\beta_i},\Delta_{\beta_{32}},x_i)
\nonumber  \label{4pointsBt}
\end{eqnarray}
\end{itemize}
Using the fusion transformation (\ref{transfost}), the equivalence
of the factorisation in the two channels can be rewritten:
\begin{equation}
\int_{\Sp}d\beta_{21}C_{\beta_{3},\beta_{21}}^{(\sigma_{4}\sigma_{3}\sigma_{1})\beta_{4}}
C_{\beta_{2}\beta_{1}}^{(\sigma_{3}\sigma_{2}\sigma_{1})\beta_{21}}%
F_{\beta_{21}\beta_{32}}\left[
\begin{array}
[c]{cc}%
\beta_{3} & \beta_{2}\\
\beta_{4} & \beta_{1}%
\end{array}
\right]
 =%
C_{\beta_{32},\beta_{1}}^{(\sigma_{4}\sigma_{2}\sigma_{1})\beta_{4}}
C_{\beta_{3}\beta_{2}}^{(\sigma_{4}\sigma_{3}\sigma_{2})\beta_{32}}
\label{pentagone boundary}
\end{equation}
The following ansatz
\begin{equation}
C_{\beta_{2}\beta_{1}}^{(\sigma_{3}\sigma_{2}\sigma_{1})\beta_{3}}=%
\frac{g(\beta_{3},\sigma_{3},\sigma_{1})}{g(\beta_{2},\sigma_{3},\sigma_{2})g(\beta_{1},\sigma_{2},\sigma_{1})}F_{\sigma_{2}\beta_{3}}\left[
\begin{array}
[c]{cc}%
\beta_{2}     & \beta_{1}\\
\sigma_{3}    &   \sigma_{1}%
\end{array}
\right] \label{definition de g}
\end{equation}
yields a solution to (\ref{pentagone boundary}), as was noticed in
\cite{R} (one recognizes the pentagonal equation
(\ref{pentagone})). The functions $g(\beta,\sigma_{2},\sigma_{1})$
appearing are unrestricted by (\ref{pentagone boundary}), and will
turn out to correspond to the normalization of the boundary
operators $B_{\beta}^{\sigma_{2}\sigma_{1}}$. We will show that it
is possible to compute this normalization explicitly, and that the
ansatz (\ref{definition de g}) is indeed consistent with the
normalization required for the boundary two point function.

\subsection{Normalization of the boundary operator}
The normalization is easily computed using the special
value\footnote{In \cite{PT3}, it is the value $\beta=-b$ that is
used to determine to normalization of the boundary operators.}
$\beta=-b/2$. The degenerate operator
$B_{-b/2}^{\sigma_{2}\sigma_{1}}$ is a {\it non vanishing} primary
field {\it in general}: it is shown in \cite{FZZ} that already at
the classical level we have
$$
\left(\frac{d^2}{dx^2}+T_{\mathrm{cl}}\right)e^{-\varphi/4}_{s,s}=\pi
b^2(\pi\mu_B^2b^2-\mu)e^{3\varphi/4}_{s,s}\, ,
$$
where $T_{\mathrm{cl}}$ is the boundary value of the classical
stress-energy tensor:
$$
T_{\mathrm{cl}}=-\frac{1}{16}\varphi^2_x+\frac{1}{4}\varphi_{xx}
+\pi b^2(\pi\mu_B^2b^2-\mu)e^{\varphi}\, ,
$$
and $\varphi=2b\phi$ is the classical Liouville field. This
relation means that generically, the second order differential
equation has some non zero terms in the right hand side, unless
$\pi \mu_B^2b^2/\mu=1$ This effect holds at the quantum level too:
if the boundary and bulk cosmological constant are related as
$$
1=\frac{2\mu^2_B}{\mu}\tan \frac{\pi b^2}{2},
$$
then the second order differential equations holds for the boundary operator
 $B_{-b/2}^{s,s}$.
It is argued in \cite{FZZ} that this remark with the structure of
singularities of the boundary two point function suggests that the
level 2 degenerate boundary operator
$B_{-b/2}^{\sigma_{2}\sigma_{1}}$ has a vanishing null vector if
and only if the right and left boundary conditions are related by
$\sigma_2=\sigma_1 \pm b/2$. Under this suggestion, the operator
product expansion of this degenerate operator with any primary
field $B_{\beta_2}^{\sigma_{3}\sigma_1}$ is truncated
\begin{eqnarray}
B_{\beta_2}^{\sigma_{3}\sigma_{2}}B_{-b/2}^{\sigma_{2}\sigma_{1}}
= c_{+}^{\pm}B_{\beta_2-b/2}^{\sigma_{3}\sigma_{1}}+c_{-}^{\pm}
B^{\sigma_{3}\sigma_{1}}_{\beta_2+b/2}\; , \quad \sigma_2=\sigma_1
\pm b/2  \label{OPE -b/2}
\end{eqnarray}
where as in the bulk situation, $c_{\pm}^{\pm}$ are structure
constants and are obtained as certain screening integrals. In the
first term of (\ref{OPE -b/2}) there is no need of screening
insertion and therefore $c_{+}^{\pm}$ can be set to 1. Let us
insert in (\ref{OPE -b/2}) the operator
$B_{Q-\beta_2+b/2}^{\sigma_{1}\sigma_{3}}$. Together with the two
point function
\begin{eqnarray}
\left\langle
B_{\beta}^{\sigma_{1}\sigma_{2}}B_{Q-\beta}^{\sigma_{2}\sigma_{1}}\right\rangle
= 1 \, ,\label{2pointsB}\nonumber
\end{eqnarray}
one gets the three point function
\begin{eqnarray}
\left\langle
B_{\beta_{2}}^{\sigma_{3}\sigma_{2}}B_{-b/2}^{\sigma_{2}\sigma_{1}}B_{Q-\beta_2+b/2}^{\sigma_{1}\sigma_{3}}\right\rangle
=1\; , \quad \sigma_2=\sigma_1 \pm b/2\, . \nonumber
\end{eqnarray}
Plugging these special values in (\ref{definition de g}), one
finds that the normalization satisfies the following first order
difference equation
\begin{equation}
1=\frac{g(\beta_{2}-b/2,\sigma_{3},\sigma_{1})}{g(\beta_{2},\sigma_{3},\sigma_{1}\pm
b/2 )g(-b/2,\sigma_{1}\pm b/2 ,\sigma_{1})}F_{\sigma_{1}\pm
b/2,\beta_{2}-b/2}\left[
\begin{array}
[c]{cc}%
\beta_{2}     & -b/2\\
\sigma_{3}    &   \sigma_{1}%
\end{array}
\right] \nonumber
\end{equation}
where the fusion coefficient on the right hand side $F_{\pm,+}$ is
explicitly written in the Appendix A. One finds for the
normalization
\begin{eqnarray}
\lefteqn{g(\beta,\sigma_3,\sigma_1)=} \nonumber \\
& & \frac{f(\sigma_3,\sigma_1)^{b^{-1}\beta
/2}\Gamma_{b}(Q)\Gamma_{b}(Q-2\beta)
\Gamma_{b}(2\sigma_{1})\Gamma_{b}(2Q-2\sigma_{3})}{\Gamma_{b}(2Q-\beta-\sigma_{1}-\sigma_{3})
\Gamma_{b}(\sigma_{1}+\sigma_{3}-\beta)\Gamma_{b}(Q-\beta+\sigma_{1}-\sigma_{3})
\Gamma_{b}(Q-\beta+\sigma_{3}-\sigma_{1})} \nonumber
\end{eqnarray}
the remaining freedom being parameterized by the function
$f(\sigma_3,\sigma_1) $. Let us furthermore note that one may
derive a second finite difference equation for the normalization
$g(\beta,\sigma_2,\sigma_1)$ obtained by replacing $b$ by
$b^{-1}$, if one considers the dual operator
$B_{-1/2b}^{\sigma_{2}\sigma_{1}}$. Taken together, these
functional relations allow one to conclude that our solution is
unique at least for irrational values of $b$. Let us now turn to
the determination of the function $f(\sigma_3,\sigma_1)$: we now
insert the operator $B_{Q-\beta_2-b/2}^{\sigma_{1}\sigma_{3}}$ and
take the expectation value
\begin{eqnarray}
\left\langle
B_{\beta_{2}}^{\sigma_{3}\sigma_{2}}B_{-b/2}^{\sigma_{2}\sigma_{1}}
B_{Q-\beta_2-b/2}^{\sigma_{1}\sigma_{3}}\right\rangle =
c_{-}^{\pm} \; .\nonumber
\end{eqnarray}
On the one hand, $c_{-}^{\pm}$ requires one insertion of the
boundary interaction $-\mu_B\int e^{b\phi(x)} dx$ and was
explicitly computed in \cite{FZZ} with the following result
\begin{eqnarray}
\lefteqn{c_{-}^\pm=
2\left(-\frac{\mu}{\pi\gamma(-b^2)}\right)^{1/2}
 \Gamma(2b\beta_2 -b^2-1)\Gamma(1-2b\beta_2)\times} \nonumber \\
&& \sin\pi b(\beta_2\pm(\sigma_1-\sigma_3) -b/2) \sin\pi b(\beta_2
\pm(\sigma_3+\sigma_1-Q)-b/2), \nonumber
\end{eqnarray}
on the other hand, (\ref{definition de g}) reads
\begin{equation}
c_{-}^{\pm}=\frac{g(\beta_{2}+b/2,\sigma_{3},\sigma_{1})}
{g(\beta_{2},\sigma_{3},\sigma_{1}\pm b/2)g(-b/2,\sigma_{1}\pm b/2
,\sigma_{1})}F_{\sigma_{1}\pm b/2,\beta_{2}+b/2}\left[
\begin{array}
[c]{cc}%
\beta_{2}     & -b/2\\
\sigma_{3}    & \sigma_{1}%
\end{array}
\right] \; ,\label{lala}\nonumber
\end{equation}
where the explicit expression for $F_{\pm,-}$ can be found in the
Appendix A. By identifying the expressions obtained, one finds
\begin{equation}
f(\sigma_3,\sigma_1)= \pi \mu \gamma(b^{2})b^{2-2b^{2}}\, . \nonumber
\end{equation}
One now collects the pieces together and the expression for the
structure constant follows \cite{PT3}
\begin{eqnarray}
\lefteqn{C_{\beta_{2}\beta_{1}}^{(\sigma_{3}\sigma_{2}\sigma_{1})\beta_{3}}
=
 \bigl(\pi \mu \gamma(b^2) b^{2-2b^2}\bigr)^{\frac{1}{2b}(\beta_3-\beta_2-\beta_1)}} \nonumber \\
&&
\times\frac{\Gamma_b(2Q-\beta_1-\beta_2-\beta_3)\Gamma_b(\beta_2+\beta_3-\beta_1)
 \Gamma_b(Q+\beta_2-\beta_1-\beta_3)\Gamma_b(Q+\beta_3-\beta_1-\beta_2)}
{\Gamma_b(2\beta_3-Q)\Gamma_b(Q-2\beta_2)\Gamma_b(Q-2\beta_1)\Gamma_b(Q)}
\nonumber \\
&& \quad\times\frac{S_b(\beta_3+\sigma_1-\sigma_3)S_b(Q+\beta_3-\sigma_3-\sigma_1)}{S_b(\beta_2+\sigma_2-\sigma_3)S_b(Q+\beta_2-\sigma_3-\sigma_2)} \nonumber \\
&& \quad \times \frac{1}{i}\int\limits_{-i\infty}^{i\infty}ds \;\;
\frac{S_b(U_1+s)S_b(U_2+s)S_b(U_3+s)S_b(U_4+s)}
{S_b(V_1+s)S_b(V_2+s)S_b(V_3+s)S_b(Q+s)} \nonumber \label{formule
fonction 3 points}
\end{eqnarray}
the coefficients $U_i$, $V_i$ and $i=1,\ldots,4$ read
$$
\begin{array}{ll}
 U_1 =\sigma_1+\sigma_2-\beta_1            &  V_1 = Q+\sigma_2-\sigma_3-\beta_1+\beta_3 \\
 U_2 = Q-\sigma_1+\sigma_{2}-\beta_1       &  V_2 = 2Q+\sigma_2-\sigma_3-\beta_1-\beta_3 \\
 U_3 = \beta_2+\sigma_2-\sigma_3           &  V_3 = 2\sigma_2 \\
 U_4 = Q-\beta_2+\sigma_2-\sigma_3         \\
\end{array}
$$

\subsection{Remarks}
\begin{enumerate}
   \item
One may explicitly check \cite{PT3} that
$$
\lim_{\beta_1 \to
0}C_{\beta_{2}\beta_{1}}^{(\sigma_{3}\sigma_{1}\sigma_{1})\beta_{3}}=\delta(\beta_3-\beta_2)+S(\beta_3,\sigma_3,
\sigma_1)\delta(\beta_3+\beta_2-Q)
$$
  \item
It can be proven that the boundary three point function is cyclic
invariant with the help of symmetry properties of the fusion
coefficients, see Appendix B.2 of \cite{PT3}.
   \item
One recovers the expression for the boundary reflection amplitude
from the boundary three point function the same way as in
\cite{ZZ} where the bulk reflection amplitude is recovered from
the bulk three point function (see equation (\ref{ral})). Using
the fact that the fusion matrix depends on conformal weights only,
and is thus invariant when $\beta_{i} \rightarrow Q-\beta_{i}$,
\begin{equation}
C_{\beta_{2}\beta_{1}}^{(\sigma_{3}\sigma_{2}\sigma_{1})Q-\beta_{3}} =%
\frac{g(Q-\beta_{3},\sigma_{3},\sigma_{1})}{g(\beta_{3},\sigma_{3},\sigma_{1})}
C_{\beta_{2}\beta_{1}}^{(\sigma_{3}\sigma_{2}\sigma_{1})\beta_{3}}\nonumber
\end{equation}
From the expression of the normalization of the boundary operator,
one recovers the boundary reflection amplitude computed in
\cite{FZZ}
\begin{eqnarray}
\lefteqn{\frac{g(Q-\beta_{3},\sigma_{3},\sigma_{1})}{g(\beta_{3},\sigma_{3},\sigma_{1})}
 = S(\beta_{3},\sigma_{3},\sigma_{1})
 =(\pi\mu\gamma(b^{2})b^{2-2b^{2}})^{\frac{1}{2b}(Q-2\beta)}\times}
\nonumber \\
& & \times
\frac{\Gamma_{b}(2\beta_{3}-Q)}{\Gamma_{b}(Q-2\beta_{3})}
\frac{S_b(\sigma_{3}+\sigma_{1}-\beta_{3})S_b(2Q-\beta_{3}-\sigma_{1}-\sigma_{3})}{S_b(\beta_{3}+\sigma_{3}-\sigma_{1})
S_{b}(\beta_{3}+\sigma_{1}-\sigma_{3})} \nonumber \label{fonction
2 points boundary}
\end{eqnarray}
    \item
    The expression of the boundary three point function in terms
    of the fusion matrix confirms the remark made in \cite{FZZ}
    that any degenerate field $B_{-nb/2}^{\sigma_{2}\sigma_{1}},\; n\in \mathbb{N}$
    has truncated operator product expansion if $\sigma_{2}=\sigma_{1}-\frac{sb}{2},\;
     s=-n,-n+2,\dots,n-2,n,$
     and are therefore in analogy with the fusion rules for
    the degenerate bulk fields (and respectively with $b$ replaced by $1/b$).
\end{enumerate}

\section*{Acknowledgements}
The author gives heartfelt thanks to the organizers of the conference, in
particular to Sasha Belavin, for the invitation to participate.  A.~Butscher
is acknowledged for improving the English of this manuscript.

\appendix

\section*{Appendix A. Some
residues of the Liouville fusion matrix}

It is well known that in the case where one of $\al_1,
\dots,\al_4$, say $\al_i$ equals $-\frac{n}{2}b-\frac{m}{2}b^{-1}$
where $n,m \in \mathbb{N}$ and where
a triple $(\Delta_{\al_4},\Delta_{\al_3},\Delta_{\al_{21}})$, $(\Delta_{\al_{21}},\Delta_{\al_2},\Delta_{\al_{1}})$
 which contains $\Delta_{\al_i}$ satisfies the fusion rules of \cite{BPZ,FF},
 one will find that the fusion coefficients that multiply the conformal blocks are residues of the
  general fusion coefficient.\\
In the case where $\al_2=-\frac{1}{2}b$, the fusion rules are:
$$\left\lbrace
\begin{array}{l}
\alpha_{21} = \alpha_{1} -s\frac{b}{2} \, ,\\
\alpha_{32} = \alpha_{3} -s'\frac{b}{2} \, , \quad s,s' =\pm \; .\\
\end{array}\right.$$
There are four entries for the fusion matrix in this special case
$$
\fus{\al_1}{-b/2}{\al_3}{\al_4}{\al_1-sb/2,}{\al_3-s'b/2} \equiv
F^{L}_{s,s'}
$$
which expressions are well known to be:
\begin{eqnarray}
F_{++}&=&\frac{\Gamma(b(2\al_1-b))\Gamma(b(b-2\al_3)+1)}
{\Gamma(b(\al_1-\al_3-\al_4+b/2)+1)\Gamma(b(\al_1-\al_3+\al_4-b/2))} \nonumber \\
F_{+-}&=& \frac{\Gamma(b(2\al_1-b))\Gamma(b(2\al_3-b)-1)}
{\Gamma(b(\al_1+\al_3+\al_4-3b/2)-1)\Gamma(b(\al_1+\al_3-\al_4-b/2))} \nonumber \\
F_{-+}&=&\frac{\Gamma(2-b(2\al_1-b))\Gamma(b(b-2\al_3)+1)}
{\Gamma(2-b(\al_1+\al_3+\al_4-3b/2))\Gamma(1-b(\al_1+\al_3-\al_4-b/2))} \nonumber \\
F_{--}&=& \frac{\Gamma(2-b(2\al_1-b))\Gamma(b(2\al_3-b)-1)}
{\Gamma(b(-\al_1+\al_3+\al_4-b/2))\Gamma(b(-\al_1+\al_3-\al_4+b/2)+1)} \nonumber \\
\label{f1/2} \nonumber
\end{eqnarray}
The dual case where $\al_2=-b^{-1}/2$ is obtained by substituting
$b$ by $b^{-1}$.

\section*{Appendix B. Special functions}
\begin{itemize}
\item{$\Ga(x)$ function}\\
The Double Gamma function introduced by Barnes \cite{Barnes} is
defined by:
\begin{eqnarray}
\nonumber \\
&&
\text{log}\Gamma_{2}(s|\omega_1,\omega_2)=\left(\frac{\partial}{\partial
t}
\sum_{n_1,n_2=0}^{\infty}(s+n_1\omega_1+n_2\omega_2)^{-t}\right)_{t=0}
\nonumber
\end{eqnarray}
Definition: $\Gamma_b(x) \equiv \frac{\Gamma_2(x|b,b^{-1})}{\Gamma_2(Q/2|b,b^{-1})}$.\\
Functional relations:
\begin{eqnarray}
\nonumber \\
&&\Ga(x+b)= \frac{\sqrt{2\pi}b^{bx-\frac{1}{2}}}{\Gamma(bx)}\Ga(x), \nonumber \\
&&\Ga(x+1/b)=
\frac{\sqrt{2\pi}b^{-\frac{x}{b}+\frac{1}{2}}}{\Gamma(x/b)}\Ga(x).
\nonumber
\end{eqnarray}
$\Ga(x)$ is a meromorphic function of $x$, whose poles are located
at
$x=-nb-mb^{-1}, n,m \in \mathbb{N}$.\\
Integral representation convergent for $0<\mathrm{Re}x$
\begin{eqnarray}
&&\text{log}\Ga(x)=\int_{0}^{\infty}\frac{dt}{t}\left\lbrack\frac{e^{-xt}-e^{-Qt/2}}{(1-e^{-bt})(1-e^{-t/b})}
-\frac{(Q/2-x)^{2}}{2}e^{-t}-\frac{Q/2-x}{t}\right\rbrack\nonumber
\end{eqnarray}
\item{$S_b(x)$ function}\\
Definition: $S_b(x)\equiv \frac{\Ga(x)}{\Ga(Q-x)}$ \nonumber\\
Functional relations:
\begin{eqnarray}
&& S_b(x+b) = 2\text{sin}(\pi bx)S_b(x),  \nonumber \\
&& S_b(x+1/b) = 2\text{sin}(\pi x/b)S_b(x). \nonumber
\end{eqnarray}
$S_b(x)$ is a meromorphic function of $x$, whose poles are located
at $x=-nb-mb^{-1}, n,m \in \mathbb{N}$,
and whose zeros are located at $x=Q+nb+mb^{-1}, n,m \in \mathbb{N}$.\\
Integral representation convergent in the strip $0<\mathrm{Re}x<Q$
\begin{eqnarray}
&&\text{log}S_b(x)=\int_{0}^{\infty}\frac{dt}{t}\left\lbrack\frac{\text{sinh}(\frac{Q}{2}-x)t}
{2\text{sinh}(\frac{bt}{2})\text{sinh}(\frac{t}{2b})}-
\frac{(Q-2x)}{t}\right\rbrack\nonumber
\end{eqnarray}
\item{$\up(x)$ function}\\
Definition: $\up(x)^{-1} \equiv \Ga(x)\Ga(Q-x)$ \nonumber \\
Functional relations:
\begin{eqnarray}
&&\up(x+b)=\frac{\Gamma(bx)}{\Gamma(1-bx)}b^{1-2bx}\up(x),\nonumber \\
&&\up(x+1/b)=\frac{\Gamma(x/b)}{\Gamma(1-x/b)}b^{2x/b-1}\up(x).
\nonumber
\end{eqnarray}
$\up(x)$ is an entire function of $x$ whose zeros are located at $x=-nb-mb^{-1}$ and $x=Q+nb+mb^{-1}$,
 $n,m \in \mathbb{N}$.\\
Integral representation convergent in the strip $0<\mathrm{Re}x<Q$
\begin{eqnarray}
&&\text{log}\up(x)=\int_{0}^{\infty}\frac{dt}{t}\left\lbrack\left(\frac{Q}{2}-x\right)^{2}e^{-t}-
\frac{\text{sinh}^{2}(\frac{Q}{2}-x)\frac{t}{2}}{\text{sinh}\frac{bt}{2}\text{sinh}\frac{t}{2b}}
\right\rbrack \nonumber
\end{eqnarray}
\end{itemize}

\section*{Appendix C. $b$-deformed hypergeometric function}
 The results
presented here come mainly from \cite{Ueno}. The
$b$-hypergeometric function $F_b(\al,\beta;\gamma;-ix)$ is
solution of the second order finite difference equation:
\begin{equation}
\left([\del+\al]_{b}[\del+\beta]_{b}-e^{-2\pi
bx}[\del]_{b}[\del+\gamma-Q]_{b}\right)F_b(\al,\beta;\gamma;-ix)=0
\label{equ. q hyp} \nonumber
\end{equation}
where $\delta_x=\frac{1}{2\pi}\partial_x,\; [x]_{b}=\frac{\text{sin}(\pi bx)}{\text{sin}\pi b^{2}}$.\\
Remark: In the following limits, the equation becomes:
$$\left\lbrace
\begin{array}{l}
b \rightarrow 0, \\
z=-e^{2\pi bx}, \\
\al = bA, \beta = bB, \gamma = bC.
\end{array}\right.$$

$$
z(1-z)\frac{d^{2}u}{dz^{2}} + [C-(A+B+1)z]\frac{du}{dz}-ABu =
0.\nonumber
$$
which is the usual hypergeometric equation. Note that in the
deformed case, there is no singularity at $z=1$ anymore.

Integral representations:
\begin{itemize}
\item
Analogue of the Barnes representation:
\begin{eqnarray}
F_b(\al,\beta;\gamma;-ix)=\frac{1}{i}\frac{S_b(\gamma)}{S_b(\al)S_b(\beta)}
\int_{-i\infty}^{+i\infty}dse^{2\pi
sx}\frac{S_b(\al+s)S_b(\beta+s)}{S_b(\gamma+s)S_b(Q+s)}\nonumber
\end{eqnarray}
The integration contour is located to the right of the poles:
$$
\left\lbrace
\begin{array}{l}
s =-\al -nb-mb^{-1}\, , \\
s =-\beta -nb-mb^{-1}\, . \\
\end{array}
\right.
$$
and to the left of the poles:
$$
\left\lbrace
\begin{array}{l}
s =Q-\gamma+nb+mb^{-1}\, , \\
s =nb+mb^{-1}\, . \\
\end{array}
\right.
$$
where $n,m \in \mathbb{N}$.\\
The integral is uniformly convergent in the set of $x \in
\mathbb{C}$ such that  $|\mathrm{Im}x|<\frac{1}{2}
\mathrm{Re}(Q+\gamma-\beta-\alpha)$.
\item
Analogue of the Euler representation:\\
Let $G_b(x)=e^{\frac{i\pi}{2}x(x-Q)}S_b(x)$.
\begin{eqnarray}
F_b(\al,\beta;\gamma;-ix)=\frac{1}{i}\frac{G_b(\gamma)}{G_b(\beta)G_b(\gamma-\beta)}
\int_{-i\infty}^{+i\infty}dse^{2i\pi
s\beta}\frac{G_b(s-ix')G_b(s+\gamma-\beta)}{G_b(s-ix'+\al)G_b(s+Q)}
\nonumber
\end{eqnarray}
where $ix'=ix+\frac{1}{2}(\al+\beta-\gamma-Q).$\\
The integration contour is located to the right of the poles:
$$
\left\lbrace
\begin{array}{l}
s =ix'-nb-mb^{-1}\, , \\
s =\beta-\gamma-nb-mb^{-1}\, . \\
\end{array}
\right.
$$
and to the left of the poles:
$$
\left\lbrace
\begin{array}{l}
s =nb+mb^{-1} \\
s =ix'-\alpha+Q+nb+mb^{-1} \\
\end{array}
\right.
$$
with $n,m \in \mathbb{N}$.
\end{itemize}

\underline{A useful transformation:}
\begin{eqnarray}
F_{b}(\al,\beta;\gamma;-ix)= e^{\pi
x(\gamma-\al-\beta)}\frac{S_b(-ix
+\frac{\gamma-\al-\beta+Q}{2})}{S_b(-ix
+\frac{-\gamma+\al+\beta+Q}{2})}F_{b}(\gamma
-\al,\gamma-\beta;\gamma;-ix) \nonumber
\end{eqnarray}

This is the equivalent in the deformed case of the classical
transformation:
\begin{equation}
{}_2F_1(a,b;c;z) = (1-z)^{c-a-b}{}_2F_{1}(c-a,c-b;c,z).\nonumber
\end{equation}

\underline{Two identities:}
\begin{eqnarray}
&&F_{b}(0,\beta;\gamma;-ix) =1,  \nonumber\\
\label{simple1} &&F_{b}(-b,\beta;\gamma;-ix)= 1+e^{2\pi
bx}\frac{\text{sin}\pi b\beta}{\text{sin}\pi b\gamma}.\nonumber
\label{simple2}
\end{eqnarray}


\begin{thebibliography}{99}

\bibitem{Polyakov} A.~Polyakov, Phys.~Lett.~\textbf{B103} (1981)
207

\bibitem{curtright}T.~Curtright, C.~Thorn, "Conformally invariant quantization of Liouville
theory" Phys.~Rev.~Lett.~\textbf{48} (1982) 1309

\bibitem{DJ} E.~D'Hoker and R.~Jackiw, ``Liouville field theory'', Phys.~Rev.~\textbf{D26} (1982) 3517.

\bibitem{DJ2} E.~D'Hoker and R.~Jackiw, ``Space translation breaking and compactification in the Liouville theory'', Phys.~Rev.~Lett.~\textbf{50} (1983) 1719-1722.

\bibitem{DFJ} E.~D'Hoker, D.Z.~Freedman and R.~Jackiw , ``$SO(2,1)$ invariant quantization of the Liouville theory'', Phys.~Rev.~\textbf{D28} (1983) 2583.


\bibitem{GN} J.-L.~Gervais, A.~Neveu, Nucl.~Phys.~\textbf{B224} (1982) 329; Nucl.~Phys.~\textbf{B238} (1984)
3125; Nucl.~Phys.~\textbf{B238} (1984) 396;
Nucl.~Phys.~\textbf{B257}[FS14] (1985) 59


\bibitem{Kazakov} V.~Kazakov, Phys.~Lett.~\textbf{B150} (1985)
282; F.~David, Nucl.~Phys.~\textbf{B257} (1985) 45;  V.~Kazakov,
I.~Kostov and A.~Migdal, Phys.~Lett.~\textbf{B157} (1985)


\bibitem{Malda} O.~Aharony, S.S.~Gubser, J.~Maldacena, H.~Ooguri
and Y.~Oz, "Large N field theories, string theory and gravity",
Phys.~Rept.~\textbf{323} (2000) 183, hep-th/9905111


\bibitem {PT1}B.~Ponsot, J.~Teschner, ``Liouville bootstrap via harmonic analysis on a noncompact quantum group.'',
 hep-th/9911110

\bibitem {PT2}B.~Ponsot, J.~Teschner, ``Clebsch-Gordan and Racah-Wigner coefficients for a
continuous series of representations of $\uq$'',  Comm.~Math.~Phys
\textbf{224} (2001) 3, math-QA/0007097



\bibitem{Takh} L.D.~Faddeev, L.A.~Takhtadjan, "Liouville model on
the lattice", Lect.~Notes in Physics \textbf{246},
Springer-Verlag, Berlin, 1986, 166-179

\bibitem{Takh2} L.A.~Takhtadjan, "Topics in quantum geometry of
Riemann surfaces: Two dimensional gravity", in: Quantum groups and
their applications in physics, Proceedings of Enrico Fermi summer
school, Como 1994. Edited by L.~Castellani and J.~Wess. IOS Press,
1996, hep-th/9409088



\bibitem{bab} O.~Babelon,  "Extended conformal algebra and Yang-Baxter equation", Phys.~Lett.~\textbf{B215}
(1988) 523-529


\bibitem {G}J.-L.~Gervais, Comm.~Math.~Phys.~\textbf{130} (1990) 252

\bibitem{GS} J.-L.~Gervais and J.~Schnittger, Phys.~Lett.~\textbf{B315} (1993) 258,  Nucl.~Phys.~\textbf{B413}
(1994) 433, Nucl.~Phys.~\textbf{B431} (1994) 273

\bibitem{CGR}E.~Cremmer, J.-L.~Gervais and J.-F.~Roussel,
Comm.~Math.~Phys.~\textbf{161} (1994) 597

\bibitem{GR} J.-L.~Gervais and J.-F.~Roussel, Nucl.~Phys.~\textbf{B426}
(1994) 140

\bibitem{Faddeev}L.D.~Faddeev, "Modular double of quantum group", math.qa/9912078





\bibitem{T}J.~Teschner, "Liouville theory revisited", Class.~Quant.~Grav. \textbf{18}: R153-R222 (2001),
 hep-th/0104158


\bibitem{R}I.~Runkel, "Boundary structure constant for the A-series Virasoro
 minimal models", Nucl.~Phys.~\textbf{B549} (1999) 563, hep-th/9811178

\bibitem {PT3}B.~Ponsot, J.~Teschner, ``Boundary Liouville Field Theory: Boundary three point function'',
  Nucl. Phys. \textbf{B622} (2002) 309, hep-th/0110244

\bibitem{BPZ}A.A.~Belavin, A.M.~Polyakov and A.B.~Zamolodchikov, ``Infinite conformal symmetry in 2D
 quantum field theory'', Nucl.~Phys.~\textbf{B241} (1984) 333


\bibitem{GL} M.~Goulian and M.~Li, ``Correlation Functions in Liouville theory'', Phys.~Rev.~Lett \textbf{66} (1991) 2051

\bibitem {FF}B.L.~Feigin, D.B.~Fuchs,
 Representation of the Virasoro algebra, in: A.~M.~Vershik, D.~P.~Zhelobenko (Eds),
  Representations of Lie groups and related topics, Gordon and Breach, London, 1990.




\bibitem{DF}V.S.~Dotsenko and V.A.~Fateev, "Four point correlations
functions and the operator algebra in the two dimensional
conformal invariant theories with the central charge $c<1$"
Nucl.~Phys.~\textbf{B251} (1985) 691




\bibitem {Dorn}H.~Dorn and H.-J.~Otto, Nucl.~Phys.~\textbf{B429} (1994) 375, hep-th/9403141


\bibitem {ZZ}A.B.~Zamolodchikov and Al.B.~Zamolodchikov, Nucl.~Phys.~\textbf{B477} (1996) 577, hep-th/9506136




\bibitem{Barnes}E.W.~Barnes, "The theory of the double
gamma function", Phil.~Trans.~Roy.~Soc. A\textbf{196} (1901) 265



\bibitem{Pol} J.~Polchinski, "Remarks on Liouville field theory"
in Strings'90, R.~Arnowitt {\it et. al.} eds, World Scientific
1991; Nucl.~Phys.~\textbf{B357} (1991) 241



\bibitem{N} A.~Neveu, unpublished






\bibitem {MS}G.~Moore, N.~Seiberg. "Classical and quantum conformal field theory", Comm.~Math.~Phys. \textbf{123}
(1989) 177

\bibitem {schmudgen2}K.~Schmuedgen, Lett.~Math.~Phys.~\textbf{37} (1996) 211



\bibitem{ZZ2} A.B.~Zamolodchikov and Al.B.~Zamolodchikov, ``Liouville field theory on a pseudosphere'', hep-th/0101152




\bibitem{FZZ}V.A. Fateev, A.B.~Zamolodchikov and Al.B.~Zamolodchikov, ``Boundary Liouville field theory I: boundary state and boundary two point function'',
hep-th/0001012




\bibitem {tesch1}J.~Teschner, "Remarks on Liouville theory with boundary", hep-th/0009138


\bibitem {T3} J.~Teschner, unpublished

\bibitem{hosomichi}K.~Hosomichi, "Bulk-Boundary Propagator in Liouville Theory on a Disc", JHEP \textbf{0111}
 044 (2001), hep-th/0108093.\\
Al.B.~Zamolodchikov, conference on Liouville field theory,
Montpellier January 1998, unpublished.













\bibitem {Ueno}M.~Nishizawa, K.~Ueno, Integral solutions of q-difference equations
 of the hypergeometric type with $|q|=1$, q-alg/9612014



\end{thebibliography}
\end{document}